\documentclass[useAMS,usenatbib]{mn2e}

 \usepackage{graphicx}

\title[Electron impact excitation of Fe XIV]{Energy levels, radiative rates and electron impact excitation rates for transitions in Fe XIV}
\author[K. M. Aggarwal and F. P. Keenan]{Kanti  M.  ~Aggarwal$^{1}$\thanks{E-mail:
 K.Aggarwal@qub.ac.uk(KMA); F.Keenan@qub.ac.uk (FPK)} and Francis  P.   ~Keenan$^{1}$ \\
$^{1}$Astrophysics Research Centre, School of Mathematics and Physics, Queen's University Belfast, Belfast BT7 1NN, Northern Ireland, UK} 
\begin{document}

\date{Accepted 2014 September 11. Received 2014  September 11; in original form 2014 July 18}

\pagerange{\pageref{firstpage}--\pageref{lastpage}} \pubyear{2014}

\maketitle

\label{firstpage}

\begin{abstract}

Energies and lifetimes are reported for the lowest 136 levels  of Fe XIV,  belonging to the (1s$^2$2s$^2$2p$^6$) 3s$^2$3p, 3s3p$^2$, 3s$^2$3d, 3p$^3$, 3s3p3d, 3p$^2$3d,  3s3d$^2$, 3p3d$^2$  and 3s$^2$4$\ell$ configurations. Additionally, radiative rates, oscillator  strengths and line strengths are calculated for all electric dipole (E1), magnetic dipole (M1), electric quadrupole (E2) and magnetic quadrupole (M2) transitions.  Theoretical lifetimes determined from these radiative rates for most  levels show satisfactory agreement  with earlier calculations, as well as with measurements.  Electron  impact excitation collision strengths are also calculated with the Dirac atomic $R$-matrix code (DARC)  over a wide energy range up to 260 Ryd. Furthermore,  resonances have been resolved in a fine energy mesh to determine  effective collision strengths, obtained after integrating the collision strengths over a Maxwellian distribution of electron velocities. Results are listed for all 9180 transitions among the 136 levels over a wide range of electron temperatures, up to  10$^{7.1}$ K. Comparisons are made with available results in the literature, and the accuracy of the  present  data is assessed.

\end{abstract}

\begin{keywords}
atomic data -- atomic processes
\end{keywords}

\section{Introduction}

Iron is an abundant element in solar and other astrophysical plasmas, and its emission lines are detected in  all ionization stages. To analyse the vast amount of observational data
available from space missions such as SOHO, {\em Chandra}, XMM-{\em Newton} and {\em Hinode}, theoretical atomic data for Fe ions are required, as there generally  is paucity of experimental results.

Emission lines of Fe XIV have been widely observed in a variety of solar and other astrophysical plasmas, and over a wide wavelength range  from the optical to extreme
ultra-violet -- see, for example, \cite{jup}, \cite{tn}, \cite{lam} and \cite{cmb}. The strongest observed solar forbidden transition is from the coronal green line (3s$^2$3p) $^2$P$^o_{1/2 }$ - $^2$P$^o_{3/2}$ at a wavelength of 5303 ${\rm {\AA}}$. This emission line has been widely used for the study of the electron density distribution in the solar corona  \citep{rrf} as well as for coronal oscillations in connection with coronal heating and energy transport  \citep{spl}. Many of the Fe XIV line pairs are density and/or temperature sensitive, and hence provide useful information about physical conditions of the plasmas (see for example, \cite{akb}, \cite{jwb} and references therein). However, to reliably analyse observations, atomic data are required for many parameters, including energy levels, radiative rates (A- values) and excitation rates. Since experimental data are generally not available, except for energy levels, theoretical results are required.  

Considering the importance of Fe XIV, particularly as a solar plasma diagnostic,  many calculations have been performed in the past, particularly for energy levels and A- values, such as those by \cite{knh}, \cite*{ps1}, \cite*{ps2},  \cite{uis}, \cite{gm1,gm2}, \cite*{cff1}, \cite{czd} and \cite{hlw}. However, most of these calculations are confined to transitions among the lowest 40 levels of the $n$ = 3 configurations, although Gupta \& Msezane  and Wei et al. have also included some levels of the $n$ = 4 configurations.  Similarly,  \cite{sst} has published energy levels, radiative rates and effective collision strengths for some transitions among the lowest 59 levels of the  (1s$^2$2s$^2$2p$^6$) 3s$^2$3p, 3s3p$^2$, 3s$^2$3d, 3p$^3$, 3s3p3d and 3p$^2$3d configurations of Fe XIV. However, in a  more recent paper, \cite{gyl} have reported A- values for a larger number of transitions among 197 levels of Fe XIV. Nevertheless, most of these workers have reported A- values for the electric dipole (E1) transitions alone, whereas in the modelling of plasmas the corresponding results are also required for other types of transitions, namely electric quadrupole (E2), magnetic dipole (M1) and magnetic quadrupole (M2).  Therefore, the aim of the present work is to extend  the earlier available calculations, and to  report a  complete set of results for  all transitions, which can be confidently applied in plasma modelling.

Corresponding calculations for electron impact excitation of Fe XIV are comparatively fewer, and the most notable ones are those by \cite{dk}, \cite{ps1,ps2}, \cite{sst} and \cite{gyl}. Most of the earlier calculations for collision strengths ($\Omega$), or more appropriately effective collision strengths ($\Upsilon$), provided conflicting estimates for the physical parameters of the plasmas -- see, for example, \cite{jwf}. Similarly, there was a large scatter between the ratios of  observed and predicted intensities for many emission lines of Fe XIV, which varied between 0.23 and 6.20, as discussed and demonstrated by \cite*{nsb}. Some of the discrepancies noted by Brickhouse et al. and \cite*{pry} were later resolved by \cite{ps2}, who also performed a large collisional calculation among the lowest 40 levels of the  3s$^2$3p, 3s3p$^2$, 3s$^2$3d, 3p$^3$ and 3s3p3d configurations. For the determination of energy levels and radiative rates, they adopted the {\em SuperStructure} (SS) code of \cite*{ss}, and for the scattering process the $R$- matrix code of \cite{rm1} was used. The calculations were basically performed in the $LS$ coupling (Russell-Saunders or spin-orbit coupling), but mass and Darwin relativistic energy shifts were included through the {\em term coupling coefficients} while transforming the $LS$ coupling reactance matrices to the intermediate coupling scheme. They calculated values of $\Omega$ up to an energy of 100 Ryd, and resolved resonances in a fine energy mesh in the thresholds region. Furthermore, they reported values of effective collision strengths ($\Upsilon$) over a wide range of electron temperatures up to 10$^{7.2}$ K, but only for transitions from the lowest two levels to the excited levels of the 3s3p$^2$ and 3s$^2$3d configurations. Therefore, their results are too limited for a full and complete application to plasma modelling. In addition, their range of partial waves ($L \le$ 18)  is generally insufficient to obtain converged values of $\Omega$, although they included the contribution of higher neglected partial waves through a top-up. Similarly, the energy range of their calculations is insufficient to provide converged values of $\Upsilon$ up T$_e$ = 10$^{7.2}$ K, although it may not affect the limited set of transitions for which they reported  results.

The limitations of  \cite{ps2} calculations were addressed by \cite{sst}, who included 135 levels of the  3s$^2$3p, 3s3p$^2$, 3s$^2$3d, 3p$^3$, 3s3p3d, 3p$^2$3d, 3s3d$^2$, 3p3d$^2$, 3s$^2$4s, 3s$^2$4p and 3s$^2$4d configurations. For the calculation of energy levels and the A- values, he adopted the  multi-configuration Hartree-Fock (MCHF) codes of  \cite{mchf} and \cite{zff}, while for the scattering process the  semi-relativistic $R$-matrix code of   \cite*{rm2} was used. Furthermore, he calculated values of $\Omega$ up to an energy of 150 Ryd, although the range of partial waves adopted ($J \le$ 25) is insufficient at higher  energies,  for the convergence of the allowed and some of the forbidden transitions. However, he included the contribution of higher neglected partial waves through a top-up, as also undertaken by  Storey et al. Similarly, he resolved resonances in the thresholds region and calculated values of $\Upsilon$ up to  T$_e$ = 10$^{7}$ K. Therefore, his calculations should provide a clear improvement over those of   Storey et al. Unfortunately, as  noted above, he reported values of $\Upsilon$ for only some transitions among the lowest 59 levels, and hence are not fully sufficient for plasma modelling where a complete set of data  are desirable   \citep*{del04}. 

The limitation in the \cite{sst} results was removed by \cite{gyl}, who not only performed a larger calculation among 197 levels of the  3s$^2$3p, 3s3p$^2$, 3s$^2$3d, 3p$^3$, 3s3p3d, 3p$^2$3d, 3s3d$^2$, 3p3d$^2$,  3s$^2$4$\ell$,   3s3p4s, 3s3p4p and  3s3p4d configurations, but also reported results of $\Upsilon$ for  all transitions among these levels. To determine the atomic structure they adopted the {\em AutoStructure} (AS) code of \cite{as}, and for  collisional data the $R$- matrix code of \cite{rm2}. Similar to the calculations of \cite{ps2}, their $\Omega$ were primarily generated in  $LS$ coupling and corresponding results for {\em fine-structure} transitions  obtained through their intermediate coupling frame transformation (ICFT) method. However, they included a larger range of partial waves (up to $J$ = 41), resolved resonances in a fine mesh of energy ($\sim$ 0.0017 Ryd), and reported values of $\Upsilon$ up to a very high temperature of $\sim$ 4 $\times$ 10$^8$ K.  Nevertheless, they left some scope for improvement. For example, they included electron exchange only up to $J$ = 12, and for $J$ $\ge$ 13 adopted a coarse mesh of energy of $\sim$ 0.34 Ryd. More importantly, they performed calculations of $\Omega$ in a limited range of energy ($\sim$ 90 Ryd),  insufficient for the accurate determination of $\Upsilon$ up to T$_e$ = 4 $\times$ 10$^8$ K ($\sim$ 2533 Ryd), although they did take into account the high energy expansion of $\Omega$, based on the formulations suggested by \cite{bt92}. Such compromises  in the calculations may explain the large discrepancies in $\Upsilon$ noted by them with previous work. For  a large number of transitions and at all temperatures,  the \cite{gyl} results are significantly higher than those of  \cite{ps2} and  \cite{sst} -- see their fig. 4. Given the importance of Fe XIV in astrophysics,  we have therefore performed yet another calculation using  a completely different approach. 

For the generation of  wavefunctions we have adopted the GRASP (general-purpose relativistic atomic structure package) code, originally developed by \cite{grasp0} but revised by Dr. P. H. Norrington. It is a fully relativistic code, based on the $jj$ coupling scheme,  available at the website {\tt http://web.am.qub.ac.uk/DARC/}, and has been successfully applied by us (and other workers) to a wide range of ions.  Further relativistic corrections arising from the Breit interaction and QED effects have also been included. Additionally, we have used the option of {\em extended average level} (EAL), in which a weighted (proportional to 2$j$+1) trace of the Hamiltonian matrix is minimized. This produces a compromise set of orbitals describing closely lying states with  moderate accuracy. For the scattering calculations, the {\em Dirac atomic $R$-matrix code} (DARC) of P. H. Norrington and I. P. Grant has been adopted. This is a  relativistic version of the standard $R$-matrix code,   is based on the $jj$ coupling scheme, and hence includes   fine-stucture in the definition of channel coupling. Subsequently,  the size of the Hamiltonian matrix increases in a calculation, but it (generally) leads  to higher accuracy  (for $\Omega$ and subsequently 

\begin{table*}
 \centering
  \caption{Energy levels (in Ryd) of Fe XIV  and lifetimes (s).  ($a{\pm}b \equiv a{\times}$10$^{{\pm}b}$).}
\begin{tabular}{rllrrrrrrrr} \hline
Index  & \multicolumn{2}{c}{Configuration/Level}  & NIST        & GRASP1a      & GRASP1b & GRASP2  &     SST & ICFT    & $\tau$ (s)\\
 \hline
    1  & 3s$^2$3p               &   $^2$P$^o_{1/2}$  & 0.00000  &     0.00000  & 0.00000 & 0.00000 & 0.00000 & 0.00000 & ........  \\
    2  & 3s$^2$3p               &   $^2$P$^o_{3/2}$  & 0.17180  &     0.17535  & 0.17074 & 0.17067 & 0.16239 & 0.16705 & 1.692-02  \\
    3  & 3s3p$^2$               &   $^4$P$  _{1/2}$  & 2.05139  &     2.03234  & 2.03064 & 2.02900 & 2.03507 & 2.01971 & 3.139-08  \\
    4  & 3s3p$^2$               &   $^4$P$  _{3/2}$  & 2.12133  &     2.10370  & 2.09992 & 2.09815 & 2.08944 & 2.08747 & 1.675-07  \\
    5  & 3s3p$^2$               &   $^4$P$  _{5/2}$  & 2.20879  &     2.19432  & 2.18761 & 2.18561 & 2.18005 & 2.17442 & 4.429-08  \\
    6  & 3s3p$^2$               &   $^2$D$  _{3/2}$  & 2.72689  &     2.74298  & 2.73751 & 2.73105 & 2.72869 & 2.71977 & 4.145-10  \\
    7  & 3s3p$^2$               &   $^2$D$  _{5/2}$  & 2.74719  &     2.76419  & 2.75717 & 2.75061 & 2.73363 & 2.73919 & 5.269-10  \\
    8  & 3s3p$^2$               &   $^2$S$  _{1/2}$  & 3.32333  &     3.37779  & 3.37470 & 3.37569 & 3.32764 & 3.34992 & 5.101-11  \\
    9  & 3s3p$^2$               &   $^2$P$  _{1/2}$  & 3.54036  &     3.60640  & 3.60098 & 3.60187 & 3.53864 & 3.57266 & 2.673-11  \\
   10  & 3s3p$^2$               &   $^2$P$  _{3/2}$  & 3.61328  &     3.68359  & 3.67667 & 3.67745 & 3.64645 & 3.64739 & 2.295-11  \\
   11  & 3s$^2$3d               &   $^2$D$  _{3/2}$  & 4.31233  &     4.40724  & 4.39829 & 4.36965 & 4.34655 & 4.37626 & 2.133-11  \\
   12  & 3s$^2$3d               &   $^2$D$  _{5/2}$  & 4.33036  &     4.42619  & 4.41499 & 4.38642 & 4.36841 & 4.39805 & 2.341-11  \\
   13  & 3p$^3$                 &   $^2$D$^o_{3/2}$  & 5.25239  &     5.26092  & 5.25358 & 5.25271 & 5.25299 & 5.23628 & 2.147-10  \\
   14  & 3p$^3$                 &   $^2$D$^o_{5/2}$  & 5.28747  &     5.29554  & 5.28593 & 5.28494 & 5.26203 & 5.26904 & 2.548-10  \\
   15  & 3p$^3$                 &   $^4$S$^o_{3/2}$  & 5.36738  &     5.39484  & 5.38808 & 5.38792 & 5.37027 & 5.36191 & 2.715-11  \\
   16  & 3s3p($^3$P)3d          &   $^4$F$^o_{3/2}$  &          &     5.86121  & 5.85238 & 5.85283 & 5.85599 & 5.85344 & 2.872-09  \\
   17  & 3s3p($^3$P)3d          &   $^4$F$^o_{5/2}$  & 5.88668  &     5.89986  & 5.88910 & 5.88950 & 5.89250 & 5.89074 & 3.185-09  \\
   18  & 3p$^3$                 &   $^2$P$^o_{1/2}$  & 5.85316  &     5.90425  & 5.89621 & 5.89598 & 5.85859 & 5.86547 & 5.830-11  \\
   19  & 3p$^3$                 &   $^2$P$^o_{3/2}$  & 5.88140  &     5.93513  & 5.92504 & 5.92464 & 5.88606 & 5.89391 & 5.910-11  \\
   20  & 3s3p($^3$P)3d          &   $^4$F$^o_{7/2}$  & 5.94097  &     5.95608  & 5.94279 & 5.94310 & 5.94483 & 5.94453 & 3.547-09  \\
   21  & 3s3p($^3$P)3d          &   $^4$F$^o_{9/2}$  & 6.01676  &     6.03421  & 6.01785 & 6.01804 & 6.01696 & 6.01885 & 1.927-02  \\
   22  & 3s3p($^3$P)3d          &   $^4$P$^o_{5/2}$  & 6.29051  &     6.32562  & 6.31451 & 6.31655 & 6.31608 & 6.31613 & 3.335-11  \\
   23  & 3s3p($^3$P)3d          &   $^4$D$^o_{3/2}$  & 6.31200  &     6.34978  & 6.33917 & 6.34120 & 6.33546 & 6.33640 & 2.734-11  \\
   24  & 3s3p($^3$P)3d          &   $^4$D$^o_{1/2}$  & 6.32572  &     6.36637  & 6.35636 & 6.35842 & 6.34811 & 6.35006 & 2.328-11  \\
   25  & 3s3p($^3$P)3d          &   $^4$P$^o_{1/2}$  & 6.41304  &     6.44657  & 6.43439 & 6.43625 & 6.42631 & 6.42949 & 3.300-11  \\
   26  & 3s3p($^3$P)3d          &   $^4$P$^o_{3/2}$  & 6.41722  &     6.45447  & 6.44143 & 6.44333 & 6.43210 & 6.43638 & 2.838-11  \\
   27  & 3s3p($^3$P)3d          &   $^4$D$^o_{7/2}$  & 6.40979  &     6.45542  & 6.44028 & 6.44219 & 6.43175 & 6.43704 & 2.318-11  \\
   28  & 3s3p($^3$P)3d          &   $^4$D$^o_{5/2}$  & 6.41636  &     6.45847  & 6.44464 & 6.44657 & 6.43433 & 6.43956 & 2.502-11  \\
   29  & 3s3p($^3$P)3d          &   $^2$D$^o_{3/2}$  & 6.53556  &     6.59908  & 6.58702 & 6.59017 & 6.56402 & 6.57354 & 2.629-11  \\
   30  & 3s3p($^3$P)3d          &   $^2$D$^o_{5/2}$  & 6.54163  &     6.60483  & 6.59156 & 6.59466 & 6.56984 & 6.57957 & 2.700-11  \\
   31  & 3s3p($^3$P)3d          &   $^2$F$^o_{5/2}$  & 6.78862  &     6.88364  & 6.87251 & 6.87715 & 6.84338 & 6.84840 & 5.516-11  \\
   32  & 3s3p($^3$P)3d          &   $^2$F$^o_{7/2}$  & 6.92393  &     7.02305  & 7.00797 & 7.01246 & 6.97121 & 6.98042 & 5.175-11  \\
   33  & 3s3p($^3$P)3d          &   $^2$P$^o_{3/2}$  & 7.35495  &     7.48700  & 7.47685 & 7.48278 & 7.40442 & 7.44321 & 1.820-11  \\
   34  & 3s3p($^3$P)3d          &   $^2$P$^o_{1/2}$  &          &     7.56294  & 7.55022 & 7.55569 & 7.47391 & 7.51553 & 2.024-11  \\
   35  & 3s3p($^1$P)3d          &   $^2$F$^o_{7/2}$  & 7.45046  &     7.58427  & 7.57027 & 7.57062 & 7.53250 & 7.54744 & 1.823-11  \\
   36  & 3s3p($^1$P)3d          &   $^2$F$^o_{5/2}$  & 7.47787  &     7.61169  & 7.59845 & 7.59883 & 7.55632 & 7.57198 & 1.771-11  \\
   37  & 3s3p($^1$P)3d          &   $^2$P$^o_{1/2}$  & 7.65001  &     7.81047  & 7.79952 & 7.80479 & 7.70946 & 7.74820 & 1.422-11  \\
   38  & 3s3p($^1$P)3d          &   $^2$D$^o_{3/2}$  & 7.66171  &     7.82117  & 7.80858 & 7.81542 & 7.75141 & 7.76494 & 1.311-11  \\
   39  & 3s3p($^1$P)3d          &   $^2$P$^o_{3/2}$  & 7.68796  &     7.84847  & 7.83550 & 7.84181 & 7.72717 & 7.78882 & 1.341-11  \\
   40  & 3s3p($^1$P)3d          &   $^2$D$^o_{5/2}$  & 7.69544  &     7.85291  & 7.83914 & 7.84730 & 7.76433 & 7.79871 & 1.185-11  \\
   41  & 3p$^2$($^1$D)3d        &   $^2$F$  _{5/2}$  &          &     8.79290  & 8.77901 & 8.77181 & 8.76349 & 8.76535 & 1.520-10  \\
   42  & 3p$^2$($^1$D)3d        &   $^2$F$  _{7/2}$  &          &     8.85638  & 8.83992 & 8.83356 & 8.82680 & 8.82720 & 1.286-10  \\
   43  & 3p$^2$($^3$P)3d        &   $^4$F$  _{3/2}$  &          &     8.87136  & 8.86119 & 8.83652 & 8.81852 & 8.82377 & 5.523-11  \\
   44  & 3p$^2$($^3$P)3d        &   $^4$F$  _{5/2}$  & 8.84793  &     8.91446  & 8.90219 & 8.87824 & 8.86051 & 8.86553 & 5.467-11  \\
   45  & 3p$^2$($^3$P)3d        &   $^4$F$  _{7/2}$  & 8.90566  &     8.97278  & 8.95785 & 8.93461 & 8.91954 & 8.92177 & 5.394-11  \\
   46  & 3p$^2$($^3$P)3d        &   $^2$P$  _{3/2}$  &          &     9.01893  & 9.00611 & 9.00385 & 8.98134 & 8.98871 & 3.985-11  \\
   47  & 3p$^2$($^3$P)3d        &   $^4$F$  _{9/2}$  & 8.96590  &     9.03865  & 9.02061 & 8.99543 & 8.99559 & 8.98318 & 5.401-11  \\
   48  & 3p$^2$($^3$P)3d        &   $^4$D$  _{1/2}$  &          &     9.06018  & 9.04870 & 9.04527 & 9.01874 & 9.02674 & 4.297-11  \\
   49  & 3p$^2$($^3$P)3d        &   $^4$D$  _{3/2}$  & 9.07430  &     9.13000  & 9.11570 & 9.11085 & 9.08454 & 9.09281 & 4.257-11  \\
   50  & 3p$^2$($^3$P)3d        &   $^4$D$  _{5/2}$  & 9.07876  &     9.13814  & 9.12319 & 9.11668 & 9.08990 & 9.09892 & 4.452-11  \\
   51  & 3p$^2$($^3$P)3d        &   $^2$P$  _{1/2}$  &           &     9.19927  &  9.18174 &  9.17963 & 9.15468  &  9.16113 & 3.761-11  \\
   52  & 3p$^2$($^3$P)3d        &   $^4$D$  _{7/2}$  &  9.14126  &     9.20148  &  9.18329 &  9.17746 & 9.15059  &  9.15923 & 4.566-11  \\
   53  & 3p$^2$($^1$D)3d        &   $^2$G$  _{7/2}$  &           &     9.37083  &  9.35533 &  9.31571 & 9.27154  &  9.29897 & 1.430-10  \\
   54  & 3p$^2$($^1$D)3d        &   $^2$G$  _{9/2}$  &           &     9.40888  &  9.39048 &  9.35153 & 9.29434  &  9.33577 & 1.435-10  \\
   55  & 3p$^2$($^1$D)3d        &   $^2$D$  _{5/2}$  &  9.41158  &     9.54495  &  9.53033 &  9.48587 & 9.45564  &  9.46853 & 1.803-11  \\
   56  & 3p$^2$($^1$D)3d        &   $^2$D$  _{3/2}$  &  9.44965  &     9.57682  &  9.56272 &  9.51950 & 9.48529  &  9.49919 & 1.899-11  \\
   57  & 3p$^2$($^3$P)3d        &   $^4$P$  _{5/2}$  &  9.51938  &     9.65260  &  9.63568 &  9.59352 & 9.55827  &  9.57017 & 1.544-11  \\
   58  & 3p$^2$($^3$P)3d        &   $^4$P$  _{1/2}$  &  9.51534  &     9.65716  &  9.64323 &  9.59977 & 9.55699  &  9.56728 & 1.298-11  \\
 \hline                                                                                                        
\end{tabular} 
\end{table*}
\clearpage
\newpage

\setcounter{table}{0}
\begin{table*}
 \centering
  \caption{Energy levels (in Ryd) of Fe XIV  and lifetimes (s).  ($a{\pm}b \equiv a{\times}$10$^{{\pm}b}$).}
\begin{tabular}{rllrrrrrrrr} \hline
Index  & \multicolumn{2}{c}{Configuration/Level}     & NIST      & GRASP1a      & GRASP1b  &   GRASP2 &     SST  & ICFT     & $\tau$ (s)\\
 \hline
   59  & 3p$^2$($^3$P)3d        &   $^4$P$  _{3/2}$  &  9.52301  &     9.65785  &  9.64277 &  9.59996 & 9.56065  &  9.57123 & 1.401-11  \\
   60  & 3p$^2$($^1$D)3d        &   $^2$P$  _{1/2}$  &           &     9.94094  &  9.92656 &  9.91499 & 9.84485  &  9.87842 & 2.345-11  \\
   61  & 3p$^2$($^1$D)3d        &   $^2$P$  _{3/2}$  &           &     9.96981  &  9.95549 &  9.93515 & 9.85545  &  9.89434 & 2.691-11  \\
   62  & 3p$^2$($^1$D)3d        &   $^2$S$  _{1/2}$  &           &    10.04211  & 10.02659 & 10.02387 & 9.96081  &  9.99112 & 2.231-11  \\
   63  & 3p$^2$($^1$S)3d        &   $^2$D$  _{3/2}$  &  9.99024  &    10.08319  & 10.06836 & 10.03172 & 9.94919  &  9.98623 & 2.916-11  \\
   64  & 3p$^2$($^1$S)3d        &   $^2$D$  _{5/2}$  &           &    10.18284  & 10.16632 & 10.10569 & 10.02262 & 10.06287 & 3.182-11  \\
   65  & 3p$^2$($^3$P)3d        &   $^2$F$  _{5/2}$  & 10.02380  &    10.21222  & 10.19747 & 10.14776 & 10.07283 & 10.11260 & 1.475-11  \\
   66  & 3p$^2$($^3$P)3d        &   $^2$F$  _{7/2}$  & 10.06952  &    10.26216  & 10.24480 & 10.18924 & 10.11673 & 10.15905 & 1.410-11  \\
   67  & 3s3d$^2$($^3$F)        &   $^4$F$  _{3/2}$  &           &    10.33307  & 10.31496 & 10.31070 & 10.31047 & 10.31372 & 1.777-11  \\
   68  & 3s3d$^2$($^3$F)        &   $^4$F$  _{5/2}$  &           &    10.34146  & 10.32214 & 10.31785 & 10.31912 & 10.32265 & 1.805-11  \\
   69  & 3s3d$^2$($^3$F)        &   $^4$F$  _{7/2}$  & 10.26393  &    10.35302  & 10.33200 & 10.32766 & 10.33112 & 10.33504 & 1.846-11  \\
   70  & 3s3d$^2$($^3$F)        &   $^4$F$  _{9/2}$  &           &    10.36761  & 10.34435 & 10.34009 & 10.34640 & 10.35077 & 1.903-11  \\
   71  & 3s3d$^2$($^3$P)        &   $^4$P$  _{1/2}$  &           &    10.65477  & 10.63624 & 10.63790 & 10.61843 & 10.63813 & 1.422-11  \\
   72  & 3s3d$^2$($^3$P)        &   $^4$P$  _{3/2}$  &           &    10.65971  & 10.64063 & 10.64230 & 10.62345 & 10.64372 & 1.438-11  \\
   73  & 3s3d$^2$($^3$P)        &   $^4$P$  _{5/2}$  &           &    10.66748  & 10.64640 & 10.64789 & 10.61705 & 10.65235 & 1.461-11  \\
   74  & 3p$^2$($^3$P)3d        &   $^2$D$  _{5/2}$  & 10.49750  &    10.82745  & 10.81230 & 10.65921 & 10.55669 & 10.61062 & 1.183-11  \\
   75  & 3p$^2$($^3$P)3d        &   $^2$D$  _{3/2}$  &           &    10.87735  & 10.86186 & 10.70049 & 10.60227 & 10.65415 & 1.183-11  \\
   76  & 3s3d$^2$($^1$D)        &   $^2$D$  _{5/2}$  &           &    11.15225  & 11.13296 & 11.08355 & 10.99852 & 11.05061 & 1.233-11  \\
   77  & 3s3d$^2$($^1$D)        &   $^2$D$  _{3/2}$  &           &    11.15820  & 11.13972 & 11.08552 & 10.99758 & 11.05003 & 1.217-11  \\
   78  & 3s3d$^2$($^1$G)        &   $^2$G$  _{7/2}$  & 10.93356  &    11.16290  & 11.14259 & 11.07693 & 11.01250 & 11.05886 & 2.220-11  \\
   79  & 3s3d$^2$($^1$G)        &   $^2$G$  _{9/2}$  & 10.93589  &    11.16671  & 11.14541 & 11.07945 & 11.01432 & 11.06128 & 2.178-11  \\
   80  & 3s3d$^2$($^3$F)        &   $^2$F$  _{5/2}$  & 11.33444  &    11.62829  & 11.61037 & 11.54361 & 11.42111 & 11.48263 & 1.084-11  \\
   81  & 3s3d$^2$($^3$F)        &   $^2$F$  _{7/2}$  & 11.34596  &    11.63928  & 11.61899 & 11.55422 & 11.43580 & 11.49754 & 1.118-11  \\
   82  & 3s3d$^2$($^1$S)        &   $^2$S$  _{1/2}$  &           &    11.85406  & 11.83668 & 11.82613 & 11.68813 & 11.76614 & 9.918-12  \\
   83  & 3s3d$^2$($^3$p)        &   $^2$P$  _{1/2}$  &           &    12.02187  & 12.00516 & 11.86971 & 11.71686 & 11.79288 & 8.195-12  \\
   84  & 3s3d$^2$($^3$P)        &   $^2$P$  _{3/2}$  &           &    12.04117  & 12.02267 & 11.88339 & 11.72474 & 11.79945 & 8.262-12  \\
   85  & 3p3d$^2$($^3$F)        &   $^4$G$^o_{5/2}$  &           &    12.48449  & 12.46645 & 12.46736 & 12.45928 & 12.46422 & 6.816-11  \\
   86  & 3p3d$^2$($^3$P)        &   $^4$G$^o_{7/2}$  &           &    12.53096  & 12.51043 & 12.51131 & 12.50063 & 12.50901 & 6.733-11  \\
   87  & 3p3d$^2$($^3$F)        &   $^4$G$^o_{9/2}$  &           &    12.59125  & 12.56760 & 12.56844 & 12.55370 & 12.56670 & 6.604-11  \\
   88  & 3p3d$^2$($^3$F)        &   $^4$G$^o_{11/2}$ &           &    12.66904  & 12.64158 & 12.64235 & 12.61846 & 12.64033 & 6.499-11  \\
   89  & 3p3d$^2$($^1$D)        &   $^2$F$^o_{5/2}$  &           &    12.79348  & 12.77416 & 12.77527 & 12.75199 & 12.75948 & 3.854-11  \\
   90  & 3p3d$^2$($^3$F)        &   $^2$D$^o_{3/2}$  &           &    12.86491  & 12.84510 & 12.84634 & 12.83374 & 12.83546 & 2.922-11  \\   
   91  & 3p3d$^2$($^1$D)        &   $^2$F$^o_{7/2}$  &           &    12.89459  & 12.87080 & 12.87197 & 12.85061 & 12.85293 & 3.811-11  \\
   92  & 3p3d$^2$($^3$F)        &   $^2$D$^o_{5/2}$  &           &    12.92026  & 12.89695 & 12.89810 & 12.85678 & 12.88759 & 3.225-11  \\
   93  & 3p3d$^2$($^3$P)        &   $^4$D$^o_{1/2}$  &           &    12.97600  & 12.95762 & 12.95874 & 12.94450 & 12.95819 & 3.013-11  \\
   94  & 3p3d$^2$($^3$P)        &   $^2$S$^o_{1/2}$  &           &    12.99686  & 12.97734 & 12.97870 & 13.00402 & 12.98585 & 1.948-11  \\
   95  & 3p3d$^2$($^3$P)        &   $^4$D$^o_{3/2}$  &           &    13.01803  & 12.99730 & 12.99839 & 12.96792 & 12.99476 & 3.124-11  \\
   96  & 3p3d$^2$($^3$P)        &   $^4$D$^o_{5/2}$  &           &    13.06047  & 13.03778 & 13.03933 & 13.00404 & 13.03206 & 2.508-11  \\
   97  & 3p3d$^2$($^3$F)        &   $^4$F$^o_{3/2}$  &           &    13.06494  & 13.04571 & 13.04864 & 13.01722 & 13.03749 & 1.508-11  \\
   98  & 3p3d$^2$($^3$F)        &   $^4$F$^o_{7/2}$  &           &    13.07361  & 13.05035 & 13.05257 & 13.05402 & 13.04542 & 2.064-11  \\
   99  & 3p3d$^2$($^3$F)        &   $^4$F$^o_{5/2}$  &           &    13.08689  & 13.06545 & 13.06793 & 13.03258 & 13.06110 & 1.728-11  \\
  100  & 3p3d$^2$($^3$F)        &   $^4$F$^o_{9/2}$  &           &    13.11152  & 13.08744 & 13.09003 & 13.08149 & 13.07704 & 1.736-11  \\
  101  & 3p3d$^2$($^3$P)        &   $^4$D$^o_{7/2}$  &           &    13.12848  & 13.10403 & 13.10592 & 13.04618 & 13.09812 & 2.138-11  \\
  102  & 3p3d$^2$($^1$G)        &   $^2$G$^o_{7/2}$  &           &    13.18181  & 13.15906 & 13.16058 & 13.09772 & 13.13120 & 2.335-11  \\
  103  & 3p3d$^2$($^1$G)        &   $^2$G$^o_{9/2}$  &           &    13.19006  & 13.16537 & 13.16715 & 13.11278 & 13.14554 & 2.334-11  \\
  104  & 3p3d$^2$($^3$P)        &   $^4$P$^o_{3/2}$  &           &    13.19304  & 13.17196 & 13.17340 & 13.15374 & 13.17168 & 2.380-11  \\
  105  & 3p3d$^2$($^3$P)        &   $^4$P$^o_{1/2}$  &           &    13.20608  & 13.18390 & 13.18527 & 13.12434 & 13.18468 & 2.219-11  \\
  106  & 3s$^2$4s               &   $^2$S$  _{1/2}$  & 13.07686  &    13.25548  & 13.24529 & 13.14370 & 13.03032 & 13.08248 & 4.012-12  \\
  107  & 3p3d$^2$($^3$P)        &   $^4$P$^o_{5/2}$  &           &    13.25717  & 13.23365 & 13.23506 & 13.20334 & 13.23196 & 2.346-11  \\
  108  & 3p3d$^2$($^1$G)        &   $^2$H$^o_{9/2}$  &           &    13.36974  & 13.34587 & 13.34706 & ........ & 13.30292 & 4.631-11  \\
  109  & 3p3d$^2$($^3$F)        &   $^4$D$^o_{7/2}$  &           &    13.42736  & 13.40323 & 13.40707 & 13.35810 & 13.39085 & 1.218-11  \\
  110  & 3p3d$^2$($^3$F)        &   $^4$D$^o_{5/2}$  &           &    13.43131  & 13.40871 & 13.41268 & 13.35696 & 13.39153 & 1.200-11  \\
  111  & 3p3d$^2$($^3$F)        &   $^4$D$^o_{3/2}$  &           &    13.43361  & 13.41236 & 13.41637 & 13.36196 & 13.39113 & 1.180-11  \\
  112  & 3p3d$^2$($^3$F)        &   $^4$D$^o_{1/2}$  &           &    13.43819  & 13.41725 & 13.42128 & 13.36664 & 13.39412 & 1.173-11  \\
  113  & 3p3d$^2$($^1$G)        &   $^2$H$^o_{11/2}$ &           &    13.47291  & 13.44681 & 13.44787 & ........ & 13.39731 & 5.240-11  \\
  114  & 3p3d$^2$($^1$D)        &   $^2$P$^o_{3/2}$  &           &    13.54657  & 13.52522 & 13.52710 & 13.47998 & 13.50738 & 1.220-11  \\
  115  & 3p3d$^2$($^1$D)        &   $^2$P$^o_{1/2}$  &           &    13.61203  & 13.58880 & 13.58929 & 13.53450 & 13.57282 & 1.248-11  \\
  116  & 3p3d$^2$($^3$P)        &   $^4$S$^o_{3/2}$  &           &    13.76928  & 13.74736 & 13.75183 & 13.66053 & 13.70923 & 7.871-12  \\
\hline                                                                                                        
\end{tabular} 
\end{table*}
\clearpage
\newpage

\setcounter{table}{0}
\begin{table*}
 \centering
  \caption{Energy levels (in Ryd) of Fe XIV  and lifetimes (s).  ($a{\pm}b \equiv a{\times}$10$^{{\pm}b}$).}
\begin{tabular}{rllrrrrrrrr} \hline
Index  & \multicolumn{2}{c}{Configuration/Level}     & NIST      & GRASP1a      & GRASP1b  &   GRASP2 &     SST  & ICFT     & $\tau$ (s)\\
\hline
  117  & 3p3d$^2$($^3$F)        &   $^2$F$^o_{5/2}$  &           &    13.83318  & 13.81117 & 13.81521 & 13.73004 & 13.76007 & 1.509-11  \\
  118  & 3p3d$^2$($^1$G)        &   $^2$F$^o_{7/2}$  &           &    13.85692  & 13.83338 & 13.83695 & 13.73955 & 13.78209 & 1.719-11  \\
  119  & 3p3d$^2$($^1$D)        &   $^2$D$^o_{3/2}$  &           &    13.98712  & 13.96663 & 13.97205 & 13.85764 & 13.91184 & 1.121-11  \\
  120  & 3p3d$^2$($^1$D)        &   $^2$D$^o_{5/2}$  &           &    14.04146  & 14.01795 & 14.02364 & 13.90501 & 13.96299 & 1.123-11  \\ 
  121  & 3p3d$^2$($^1$S)        &   $^2$P$^o_{1/2}$  &           &    14.04675  & 14.02866 & 14.02414 & 13.94378 & 13.98707 & 2.264-11  \\
  122  & 3p3d$^2$($^1$S)        &   $^2$P$^o_{3/2}$  &           &    14.16445  & 14.14214 & 14.13714 & 14.04766 & 14.09894 & 2.233-11  \\
  123  & 3p3d$^2$($^3$F)        &   $^2$F$^o_{7/2}$  &           &    14.24202  & 14.22046 & 14.22687 & 14.08701 & 14.12216 & 1.305-11  \\
  124  & 3p3d$^2$($^1$G)        &   $^2$F$^o_{5/2}$  &           &    14.30263  & 14.27985 & 14.28582 & 14.12242 & 14.18067 & 1.396-11  \\
  125  & 3s$^2$4p               &   $^2$P$^o_{1/2}$  & 14.29632  &    14.31291  & 14.30321 & 14.18731 & 14.13387 & 14.07465 & 1.743-11  \\
  126  & 3p3d$^2$($^3$F)        &   $^2$G$^o_{9/2}$  &           &    14.35240  & 14.32774 & 14.33480 & 14.15420 & 14.21792 & 1.266-11  \\
  127  & 3s$^2$4p               &   $^2$P$^o_{3/2}$  & 14.34343  &    14.37783  & 14.36636 & 14.24844 & 14.18940 & 14.14366 & 1.802-11  \\
  128  & 3p3d$^2$($^3$F)        &   $^2$G$^o_{7/2}$  &           &    14.38156  & 14.35734 & 14.36431 & 14.16256 & 14.24911 & 1.281-11  \\
  129  & 3p3d$^2$($^3$P)        &   $^2$D$^o_{5/2}$  &           &    14.71748  & 14.69503 & 14.70360 & 14.50304 & 14.57380 & 9.667-12  \\
  130  & 3p3d$^2$($^3$P)        &   $^2$D$^o_{3/2}$  &           &    14.72401  & 14.70309 & 14.71140 & 14.51296 & 14.57864 & 9.465-12  \\
  131  & 3p3d$^2$($^3$P)        &   $^2$P$^o_{1/2}$  &           &    14.92128  & 14.90027 & 14.89664 & 14.68134 & 14.76515 & 7.765-12  \\
  132  & 3p3d$^2$($^3$P)        &   $^2$P$^o_{3/2}$  &           &    14.93990  & 14.91830 & 14.91445 & 14.69930 & 14.78444 & 7.704-12  \\
  133  & 3s$^2$4d               &   $^2$D$  _{3/2}$  & 15.45491  &    15.72443  & 15.71151 & 15.61108 & 15.48934 & 15.49803 & 3.455-12  \\
  134  & 3s$^2$4d               &   $^2$D$  _{5/2}$  & 15.46684  &    15.73484  & 15.72089 & 15.62151 & 15.50092 & 15.51137 & 3.412-12  \\
  135  & 3s$^2$4f               &   $^2$F$^o_{5/2}$  & 16.29929  &    16.53545  & 16.52042 & 16.47274 & ........ & 16.35952 & 1.211-12  \\
  136  & 3s$^2$4f               &   $^2$F$^o_{7/2}$  & 16.29692  &    16.53847  & 16.52331 & 16.47576 & ........ & 16.36331 & 1.214-12  \\
 \hline                                                                                                        
\end{tabular}      
\begin{flushleft}
{\small
NIST: {\tt http://www.nist.gov/pml/data/asd.cfm} \\
GRASP1a: Energies from the {\sc grasp} code with 136 level calculations  without Breit and QED effects \\
GRASP1b: Energies from the {\sc grasp} code with  136 level calculations with    Breit and QED effects  \\
GRASP2: Energies from the {\sc grasp} code with 332 level calculations with    Breit and QED effects   \\
SST:   Energies of  \cite{sst} from the {\sc mchf} code \\
ICFT:  Energies of  \cite{gyl} from the {\sc as} code \\
}
\end{flushleft}
\end{table*}
\clearpage
\newpage
 $\Upsilon$), especially for transitions among the {\em fine-structure} levels of a state, because resonances through the energies of degenerating levels are also taken into  account. 
 Since  degeneracy among some of the levels is significant, see for example the 3s3p($^3$P)3d $^4$F$^o_{3/2,5/2,7/2,9/2}$ levels  in Table 1, the use of this code, we believe,  is more appropriate. It  is  unpublished but is freely available at the website {\tt http://web.am.qub.ac.uk/DARC/}, and has been successfully applied by us and other users to a wide range of ions. 

\section{Energy levels}

The (1s$^2$2s$^2$2p$^6$) 3s$^2$3p, 3s3p$^2$, 3s$^2$3d, 3p$^3$, 3s3p3d, 3p$^2$3d,  3s3d$^2$, 3p3d$^2$  and 3s$^2$4$\ell$ configurations of Fe XIV give rise to the lowest 136 levels listed in Table 1, where we compare our energies with the experimental values compiled by the NIST (National Institute of Standards and Technology) team \citep{nist}, available at their website  {\tt http://www.nist.gov/pml/data/asd.cfm}. Energy levels obtained without (GRASP1a) and with (GRASP1b)  the  Breit and QED corrections are  included in the Table. The inclusion of Breit and QED effects has (in general) lowered the energies by a maximum of 0.03 Ryd, i.e.  $\le$ 0.3\%. Additionally, their inclusion has  altered the orderings in a few instances, such as for levels 26/27, 58/59 and 106/107. The energy differences for these swapped levels are very small and we have retained the original orderings because subsequent tables for  A- and $\Upsilon$ values follow these.  However, the effect of these corrections is significant for the 3s$^2$3p $^2$P$^o_{3/2}$ level (2), where the energy has become lower by 0.00461 Ryd, i.e. 2.6\%. As a result, there is now a better match for this level with the experimental energy of NIST. For a majority of levels, there is good agreement between our calculations and the experimental values, both in magnitude and orderings. Differences, if any, are within 0.3 Ryd -- see levels 37-40 or 78-81. Similarly, there are some minor differences in the level orderings, but the most notable is level 18 (3p$^3$ $^2$P$^o_{1/2}$).

The differences between our energy levels from GRASP  and those of \cite{sst} are up to 0.3 Ryd (1.5\%) for a few levels, such as 82-84 and 126-132. Furthermore, the level orderings of Tayal do not match in a few cases  with our calculations or the experimental results -- see, for example, levels 38/39. More importantly, his energy for level 2 (3s$^2$3p $^2$P$^o_{3/2}$) is  lower than the experimental or any other theoretical value listed in Table 1, by up to 5\%. This is in spite of the fact that he adopted non-orthogonal orbitals and hence optimised each state separately and independently, a procedure that generally yields better agreement between theory and measurement. However, we note here that although this level is very important as stated in section 1, a difference of 5\% is not uncommon among Fe XIV calculations -- see, for example, table 4 of \cite{hlw}. Our calculated energies are based only on spectroscopic levels, but can be improved with additional CI or the inclusion of pseudo-orbitals. However, this is not possible bearing in mind our further calculations for other important parameters, particularly the collision strengths. For this reason, the accuracy of the energy levels (wave functions) included in the collisional calculation by Tayal is lower by $\sim$1\% than those listed in Table 1, because he deleted all those configurations  with coefficients less than 0.02 in magnitude.

\cite{gyl}  adopted the {\em AutoStructure} (AS) code of \cite{as} to calculate energies for 197 levels of Fe XIV. The 61 levels in addition to our calculation have arisen from the 3d$^3$ and 3s3p4s/4p/4d configurations. In fact, these four configurations give rise to 67 levels, but they omitted 6  from  3s3p4d.  However, to improve the accuracy they included further CI with additional 77 configurations, namely 3s3p4f, 3p$^2$4$\ell$, 3p3d4$\ell$, 3d$^2$4$\ell$, 3$\ell$4$\ell$4$\ell'$ and 3$\ell$3$\ell'$5$\ell$. As a result,  their calculated energies are expected to be more accurate and are included in Table 1 for comparison. Differences with the NIST compilations are generally smaller, but are up to 1.4\% for some  levels, such as 3s$^2$4p $^2$P$^o_{1/2,3/2}$, for which our GRASP1b energies agree better with the measurements. Similarly, their calculated energy for the 3s$^2$3p $^2$P$^o_{3/2}$ level (2) is lower than the experimental value by $\sim$3\%, only slightly improving over the  \cite{sst} result. Nevertheless, differences with our calculated (GRASP1b) energies are up to 0.2 Ryd for some of the levels, such as 3s$^2$4p $^2$P$^o_{1/2,3/2}$, as noted above.

To assess the effect of additional CI, we have also performed a larger calculation (GRASP2) with 332 levels, which arise from the 3s$^2$3p, 3s3p$^2$, 3s$^2$3d, 3p$^3$, 3s3p3d, 3p$^2$3d,  3s3d$^2$, 3p3d$^2$, 3d$^3$, 3s$^2$4$\ell$, 3s3p4$\ell$, 3s$^2$5$\ell$ and 3s3p5$\ell$ configurations. Almost all the additional 196 levels of the  3d$^3$, 3s3p4$\ell$, 3s$^2$5$\ell$ and 3s3p5$\ell$ configurations yield energies {\em above} those of the 136 included in GRASP1 and listed in Table 1. Hence their inclusion in a collisional calculation may improve the accuracy of $\Upsilon$, because of the additional resonances arising from these levels, but the effect on the energies of lower levels will not be significant, because there is no intermixing among the levels. This is confirmed by the comparison shown in Table 1 as most of the levels agree within 0.05 Ryd and the orderings are also nearly the same. However, for some of the levels the differences between the GRASP1 and GRASP2 energies are up to 0.15 Ryd -- see for example, levels 74/75, 83/84 and 125/127. In some cases the energies obtained in GRASP2 have become better (i.e. are closer to those of NIST) but are worse for others, such as 125/127, i.e. 3s$^2$4p $^2$P$^o_{1/2,3/2}$. Therefore, there is no overall advantage of extensive CI for the determination of energy levels for Fe XIV. Finally, based on the comparisons shown in Table 1 and discussed henceforth we may conclude that our calculated energy levels are accurate to about 1\%. 

\begin{table*}                                                                                                                                                  
\caption{Transition wavelengths ($\lambda_{ij}$ in $\rm \AA$), radiative rates (A$_{ji}$ in s$^{-1}$), oscillator strengths (f$_{ij}$, dimensionless), and line     
strengths (S, in atomic units) for electric dipole (E1), and A$_{ji}$ for E2, M1 and M2 transitions in Fe XIV. ($a{\pm}b \equiv a{\times}$10$^{{\pm}b}$). (For complete table see Supporting Information.)}      
\begin{tabular}{rrrrrrrrr}                                                                                                                                                                                                                                                                                              
\hline                                                                                                                                                          
$i$ & $j$ & $\lambda_{ij}$ & A$^{{\rm E1}}_{ji}$  & f$^{{\rm E1}}_{ij}$ & S$^{{\rm E1}}$ & A$^{{\rm E2}}_{ji}$  & A$^{{\rm M1}}_{ji}$ & A$^{{\rm M2}}_{ji}$ \\  
\hline                                                                                                                                                          
    1 &    2 &  5.337$+$03 &  0.000$-$00 &  0.000$-$00 &  0.000$-$00 &  1.431$-$02 &  5.907$+$01 &  0.000$-$00 \\       
    1 &    3 &  4.488$+$02 &  2.313$+$07 &  6.985$-$04 &  2.064$-$03 &  0.000$-$00 &  0.000$-$00 &  0.000$-$00 \\       
    1 &    4 &  4.340$+$02 &  4.779$+$05 &  2.698$-$05 &  7.710$-$05 &  0.000$-$00 &  0.000$-$00 &  2.089$-$00 \\       
    1 &    5 &  4.166$+$02 &  0.000$-$00 &  0.000$-$00 &  0.000$-$00 &  0.000$-$00 &  0.000$-$00 &  1.357$-$00 \\       
    1 &    6 &  3.329$+$02 &  2.334$+$09 &  7.754$-$02 &  1.699$-$01 &  0.000$-$00 &  0.000$-$00 &  1.757$-$02 \\       
    1 &    7 &  3.305$+$02 &  0.000$-$00 &  0.000$-$00 &  0.000$-$00 &  0.000$-$00 &  0.000$-$00 &  6.144$-$00 \\       
    1 &    8 &  2.700$+$02 &  1.812$+$10 &  1.981$-$01 &  3.521$-$01 &  0.000$-$00 &  0.000$-$00 &  0.000$-$00 \\       
    1 &    9 &  2.531$+$02 &  1.515$+$10 &  1.454$-$01 &  2.423$-$01 &  0.000$-$00 &  0.000$-$00 &  0.000$-$00 \\       
    1 &   10 &  2.479$+$02 &  8.171$+$09 &  1.505$-$01 &  2.456$-$01 &  0.000$-$00 &  0.000$-$00 &  1.785$-$00 \\      
    .  &    .    &  . & . & . & . & . & . & .\\ 
    .  &    .    &  . & . & . & . & . & . & .\\     
        .  &    .    &  . & . & . & . & . & . & .\\ 
\hline                                                                                                                                                          
\end{tabular}                                                                                                                                                   
\end{table*}

\section{Radiative rates}

The absorption oscillator strength (f$_{ij}$) and radiative rate A$_{ji}$ (in s$^{-1}$) for a transition $i \to j$ are related by the following expression:

\begin{equation}
f_{ij} = \frac{mc}{8{\pi}^2{e^2}}{\lambda^2_{ji}} \frac{{\omega}_j}{{\omega}_i} A_{ji}
 = 1.49 \times 10^{-16} \lambda^2_{ji}    \frac{{ \omega}_j}{{\omega}_i} A_{ji}  
\end{equation}
where $m$ and $e$ are the electron mass and charge, respectively, $c$ is the velocity of light, $\lambda_{ji}$ is the transition energy/wavelength in $\rm \AA$, and $\omega_i$ and $\omega_j$ are the statistical weights of the lower ($i$) and upper ($j$) levels, respectively. Similarly, the oscillator strength f$_{ij}$ (dimensionless) and the line strength S (in atomic unit) are related by the standard equations listed below.

\begin{flushleft}
For the electric dipole (E1) transitions
\end{flushleft}
\begin{equation}
A_{ji} = \frac{2.0261\times{10^{18}}}{{{\omega}_j}\lambda^3_{ji}} S^{{\rm E1}} \hspace*{0.5 cm} {\rm and} \hspace*{0.5 cm}
f_{ij} = \frac{303.75}{\lambda_{ji}\omega_i} S^{{\rm E1}}, \\
\end{equation}
\begin{flushleft}
for the magnetic dipole (M1) transitions
\end{flushleft}
\begin{equation}
A_{ji} = \frac{2.6974\times{10^{13}}}{{{\omega}_j}\lambda^3_{ji}} S^{{\rm M1}} \hspace*{0.5 cm} {\rm and} \hspace*{0.5 cm}
f_{ij} = \frac{4.044\times{10^{-3}}}{\lambda_{ji}\omega_i} S^{{\rm M1}}, \\
\end{equation}
\begin{flushleft}
for the electric quadrupole (E2) transitions
\end{flushleft}
\begin{equation}
A_{ji} = \frac{1.1199\times{10^{18}}}{{{\omega}_j}\lambda^5_{ji}} S^{{\rm E2}} \hspace*{0.5 cm} {\rm and} \hspace*{0.5 cm}
f_{ij} = \frac{167.89}{\lambda^3_{ji}\omega_i} S^{{\rm E2}},
\end{equation}

\begin{flushleft}
and for the magnetic quadrupole (M2) transitions
\end{flushleft}
\begin{equation}
A_{ji} = \frac{1.4910\times{10^{13}}}{{{\omega}_j}\lambda^5_{ji}} S^{{\rm M2}} \hspace*{0.5 cm} {\rm and} \hspace*{0.5 cm}
f_{ij} = \frac{2.236\times{10^{-3}}}{\lambda^3_{ji}\omega_i} S^{{\rm M2}}. \\
\end{equation}
In Table 2 we present transition energies/wavelengths ($\lambda$, in $\rm \AA$), radiative rates (A$_{ji}$, in s$^{-1}$), oscillator strengths (f$_{ij}$, dimensionless), and line
strengths (S, in a.u.), in length  form only, for all 2733 electric dipole (E1) transitions among the 136 levels of Fe XIV. The  indices used  to represent
the lower and upper levels of a transition have already been defined in Table 1. Similarly, there are {3776 E2, 2791  M1 and 3643 M2 transitions among the 136 levels. 
However,  only their  A- values are listed in Table 2, as this is the quantity required for plasma modelling. The corresponding results for f-  and S- values can be easily obtained through the above equations.  Furthermore, the  S- values have been calculated in both Babushkin and Coulomb gauges, i.e.  the length and velocity forms in the widely used non-relativistic nomenclature, but  in Table 2  results are listed in the length form alone, because  these are generally  considered to be comparatively more accurate.  Nevertheless,  we will discuss  later the velocity/length  form ratio, as this provides some assessment of  the accuracy of the results.

\begin{table*}
 \centering
 \caption{Comparison of A- values (s$^{-1}$) for some transitions of Fe XIV.  ($a{\pm}b \equiv a{\times}$10$^{{\pm}b}$).}
\begin{tabular}{rrllccllll} 
\hline
\multicolumn{2}{c}{Transition} & \multicolumn{2}{c}{GRASP1} & GRASP2    & MCHF      &    AS     & R          \\
\hline
I & J  & A & f & A & A & A &  \\
\hline
     1  &  3 &  2.313$+$07 &  6.985$-$4 & 2.313$+$07  & 2.33$+$07 & 2.29$+$07 &  8.9$-$1  \\
     1  &  4 &  4.779$+$05 &  2.698$-$5 & 5.133$+$05  & 5.68$+$05 & 5.06$+$05 &  1.2$-$0  \\
     1  &  6 &  2.334$+$09 &  7.754$-$2 & 2.341$+$09  & 2.41$+$09 & 2.40$+$09 &  1.1$-$0  \\
     1  &  8 &  1.812$+$10 &  1.981$-$1 & 1.820$+$10  & 1.70$+$10 & 1.78$+$10 &  9.7$-$1  \\
     1  &  9 &  1.515$+$10 &  1.454$-$1 & 1.511$+$10  & 1.49$+$10 & 1.49$+$10 &  1.0$-$0  \\
     1  & 10 &  8.171$+$09 &  1.505$-$1 & 8.212$+$09  & 7.80$+$09 & 7.95$+$09 &  9.9$-$1  \\
     1  & 11 &  3.836$+$10 &  4.937$-$1 & 3.804$+$10  & 3.69$+$10 & 3.81$+$10 &  1.0$-$0  \\
     2  &  3 &  8.724$+$06 &  1.570$-$4 & 8.788$+$06  & 9.16$+$06 & 8.88$+$06 &  8.1$-$1  \\
     2  &  4 &  5.491$+$06 &  1.837$-$4 & 5.496$+$06  & 5.43$+$06 & 5.45$+$06 &  9.5$-$1  \\
     2  &  5 &  2.258$+$07 &  1.036$-$3 & 2.322$+$07  & 2.33$+$07 & 2.27$+$07 &  1.0$-$0  \\
     2  &  6 &  7.886$+$07 &  1.490$-$3 & 8.179$+$07  & 8.79$+$07 & 8.82$+$07 &  1.2$-$0  \\
     2  &  7 &  1.898$+$09 &  5.298$-$2 & 1.908$+$09  & 1.98$+$09 & 1.97$+$09 &  1.1$-$0  \\
     2  &  8 &  1.484$+$09 &  8.999$-$3 & 1.521$+$09  & 1.63$+$09 & 1.58$+$09 &  8.8$-$1  \\
     2  &  9 &  2.226$+$10 &  1.177$-$1 & 2.231$+$10  & 2.14$+$10 & 2.19$+$10 &  9.7$-$1  \\
     2  & 10 &  3.541$+$10 &  3.586$-$1 & 3.534$+$10  & 3.41$+$10 & 3.46$+$10 &  9.9$-$1  \\
     2  & 11 &  8.535$+$09 &  5.946$-$2 & 8.518$+$09  & 8.26$+$09 & 8.47$+$09 &  1.0$-$0  \\
     2  & 12 &  4.272$+$10 &  4.428$-$1 & 4.239$+$10  & 4.15$+$10 & 4.27$+$10 &  1.0$-$0  \\
     3  & 13 &  1.879$+$08 &  4.503$-$3 & 1.821$+$08  &           & 1.99$+$08 &  9.9$-$1  \\
     3  & 15 &  6.713$+$09 &  1.483$-$1 & 6.635$+$09  & 6.59$+$09 & 6.56$+$09 &  9.8$-$1  \\
     4  & 13 &  1.312$+$08 &  1.642$-$3 & 1.269$+$08  &           & 1.47$+$08 &  9.9$-$1  \\
     4  & 14 &  2.962$+$06 &  5.450$-$5 & 2.901$+$06  &           & 2.70$+$06 &  1.0$-$0  \\
     4  & 15 &  1.268$+$10 &  1.460$-$1 & 1.254$+$10  & 1.25$+$10 & 1.24$+$10 &  9.8$-$1  \\
     5  & 13 &  5.092$+$08 &  4.496$-$3 & 4.957$+$08  &           & 5.44$+$08 &  9.5$-$1  \\
     5  & 14 &  8.486$+$07 &  1.100$-$3 & 8.333$+$07  &           & 8.47$+$07 &  9.2$-$1  \\
     5  & 15 &  1.730$+$10 &  1.401$-$1 & 1.710$+$10  & 1.70$+$10 & 1.69$+$10 &  9.8$-$1  \\
     6  & 13 &  2.151$+$09 &  4.230$-$2 & 2.098$+$09  & 2.23$+$09 & 2.20$+$09 &  9.4$-$1  \\
     6  & 14 &  2.984$+$08 &  8.580$-$3 & 2.898$+$08  & 3.09$+$08 & 3.04$+$08 &  9.4$-$1  \\
     6  & 15 &  7.619$+$07 &  1.350$-$3 & 7.244$+$07  &           & 8.32$+$07 &  8.1$-$1  \\
     7  & 13 &  8.838$+$08 &  1.177$-$2 & 8.615$+$08  & 9.00$+$08 & 8.94$+$08 &  8.2$-$1  \\
     7  & 14 &  2.943$+$09 &  5.729$-$2 & 2.868$+$09  & 3.04$+$09 & 3.01$+$09 &  9.2$-$1  \\
     7  & 15 &  9.874$+$06 &  1.184$-$4 & 1.023$+$07  &           & 1.21$+$07 &  1.9$-$0  \\
     8  & 13 &  3.161$+$08 &  2.229$-$2 & 3.063$+$08  &           & 3.20$+$08 &  1.1$-$0  \\
     8  & 15 &  2.867$+$05 &  1.761$-$5 & 2.784$+$05  &           & 7.56$+$04 &  2.3$-$0  \\
     9  & 13 &  4.448$+$08 &  4.055$-$2 & 4.357$+$08  &           & 4.75$+$08 &  1.1$-$0  \\
     9  & 15 &  5.653$+$06 &  4.407$-$4 & 5.218$+$06  &           & 6.51$+$06 &  1.4$-$0  \\
    10  & 13 &  2.613$+$07 &  1.308$-$3 & 2.505$+$07  &           & 2.66$+$07 &  1.3$-$0  \\
    10  & 14 &  5.882$+$08 &  4.242$-$2 & 5.742$+$08  &           & 6.22$+$08 &  1.1$-$0  \\
    10  & 15 &  5.494$+$07 &  2.335$-$3 & 5.400$+$07  &           & 5.65$+$07 &  9.4$-$1  \\
    11  & 13 &  5.966$+$06 &  1.015$-$3 & 5.882$+$06  &           & 5.08$+$06 &  3.2$-$1  \\
    11  & 14 &  3.817$+$05 &  9.047$-$5 & 3.572$+$05  &           & 3.22$+$05 &  2.1$-$1  \\
    11  & 15 &  4.099$+$05 &  5.209$-$5 & 4.297$+$05  &           & 3.90$+$05 &  5.0$-$1  \\
    12  & 13 &  6.463$+$05 &  7.628$-$5 & 7.186$+$05  &           & 6.11$+$05 &  5.4$-$1  \\
    12  & 14 &  6.829$+$06 &  1.121$-$3 & 6.784$+$06  &           & 5.80$+$06 &  3.3$-$1  \\
    12  & 15 &  2.627$+$04 &  2.302$-$6 & 1.293$+$04  &           & 1.12$+$04 &  2.3$-$2  \\
 \hline                                                                                 
\end{tabular}      
\begin {flushleft}                                                                      
                                           
\begin{tabbing}
aaaaaaaaaaaaaaaaaaaaaaaaaaaaaaaaaaaa\= \kill
GRASP1: Present results for 136 levels with the {\sc grasp} code \\
GRASP2: Present results for 332 levels with the {\sc grasp} code \\
MCHF: \cite{sst} \\ 
AS: \cite{gyl} \\ 
R: Ratio of velocity/length form of f- values from the GRASP1 calculations \\
\end{tabbing}
\end {flushleft}                                                                     
\end{table*}                                                                                   

\begin{table*}
 \centering
\begin{minipage}{180mm}
 \caption{Comparison of lifetimes ($\tau$) for the lowest 40 levels of Fe XIV. All values are in ns {\em except} when specified under the column GRASP1.}
 {\small
\begin{tabular}{rlllllll} \hline
 & & & & & &  \\
Index  & \multicolumn{2}{c}{Configuration/Level}    &GRASP1   & GRASP2        &  CIV3        & MCHF      & Experimental                                                     \\
 & & & & & &  \\ \hline
 & & & & & &  \\
   1  &  3s$^2$3p           &  $^2$P$^o$$_{1/2}$    &  .....  &      .....     &   ....      &    .....  &                                                                     \\
   2  &  3s$^2$3p           &  $^2$P$^o$$_{3/2}$    & 16.93 ms  &    16.95     &   ....      &    .....  & 17.52$\pm$0.29a, 16.74$\pm$0.12b, 17.0$\pm$0.2c, 16.73$^{+0.02}_{-0.01}$d  \\
   3  &  3s3p$^2$           &  $^4$P$$$$$_{1/2}$    & 31.39   &    31.33       &   27.3898   &    27.53  & 29$\pm$3e                                                           \\
   4  &  3s3p$^2$           &  $^4$P$$$$$_{3/2}$    & 167.5   &   166.4        &  153.0643   &   155.6   &                                                                     \\
   5  &  3s3p$^2$           &  $^4$P$$$$$_{5/2}$    & 44.29   &    43.08       &   39.4166   &   40.14   & 39$\pm$5e                                                           \\
   6  &  3s3p$^2$           &  $^2$D$$$$$_{3/2}$    & 0.4145  &     0.4127     &    0.4209   &   0.3927  & 0.550$\pm$0.020, 0.460$\pm$0.050, 0.515$\pm$0.025, 0.340$\pm$0.060f \\
   7  &  3s3p$^2$           &  $^2$D$$$$$_{5/2}$    & 0.5269  &     0.5241     &    0.5414   &   0.5005  & 0.700$\pm$0.020, 0.630$\pm$0.025, 0.625$\pm$0.025, 0.530$\pm$0.040f \\
   8  &  3s3p$^2$           &  $^2$S$$$$$_{1/2}$    & 0.0510  &     0.0507     &    0.0531   &   0.0532  & 0.077$\pm$0.004, 0.054$\pm$0.004, 0.073$\pm$0.003, 0.061$\pm$0.006f \\
   9  &  3s3p$^2$           &  $^2$P$$$$$_{1/2}$    & 0.0267  &     0.0267     &    0.0280   &   0.0274  & 0.045$\pm$0.002, 0.045$\pm$0.003, 0.045$\pm$0.002, 0.035$\pm$0.007f \\
  10  &  3s3p$^2$           &  $^2$P$$$$$_{3/2}$    & 0.0229  &     0.0223     &    0.0243   &   0.0237  & 0.046$\pm$0.003, 0.030$\pm$0.003, 0.044$\pm$0.002, 0.034$\pm$0.007f \\
  11  &  3s$^2$3d           &  $^2$D$$$$$_{3/2}$    & 0.0213  &     0.0215     &    0.0223   &   .....   & 0.039$\pm$0.002, 0.038$\pm$0.002, 0.037$\pm$0.002, 0.032$\pm$0.006f \\
  12  &  3s$^2$3d           &  $^2$D$$$$$_{5/2}$    & 0.0234  &     0.0236     &    0.0245   &   .....   & 0.041$\pm$0.002, 0.040$\pm$0.002, 0.040$\pm$0.002, 0.032$\pm$0.005f \\
  13  &  3p$^3$             &  $^2$D$^o$$_{3/2}$    & 0.2147  &     0.2204     &    0.2143   &   0.1915  &                                                                     \\
  14  &  3p$^3$             &  $^2$D$^o$$_{5/2}$    & 0.2548  &     0.2614     &    0.2576   &   0.2392  &                                                                     \\
  15  &  3p$^3$             &  $^4$S$^o$$_{3/2}$    & 0.0271  &     0.0275     &    0.0287   &   0.0279  &                                                                     \\
  16  &  3s3p($^3$P)3d      &  $^4$F$^o$$_{3/2}$    & 2.8721  &     2.902      &    0.4829   &   0.5013  & 1.5$\pm$0.2g                                                        \\
  17  &  3s3p($^3$P)3d      &  $^4$F$^o$$_{5/2}$    & 3.1855  &     3.233      &    2.9590   &   3.229   & 1.9$\pm$0.1g                                                        \\
  18  &  3p$^3$             &  $^2$P$^o$$_{1/2}$    & 0.0583  &     0.0593     &    0.0590   &   0.0593  &                                                                     \\
  19  &  3p$^3$             &  $^2$P$^o$$_{3/2}$    & 0.0591  &     0.0602     &    0.0719   &   0.0739  &                                                                     \\
  20  &  3s3p($^3$P)3d      &  $^4$F$^o$$_{7/2}$    & 3.547   &     3.606      &    3.3986   &   3.718   & 2.8$\pm$0.2g                                                        \\
  21  &  3s3p($^3$P)3d      &  $^4$F$^o$$_{9/2}$    & 19.27 ms   &    19.29    &   ....      &   .....   &                                                                     \\
  22  &  3s3p($^3$P)3d      &  $^4$P$^o$$_{5/2}$    & 0.0334  &     0.0335     &    0.0347   &   0.0336  &                                                                     \\
  23  &  3s3p($^3$P)3d      &  $^4$D$^o$$_{3/2}$    & 0.0273  &     0.0275     &    0.0266   &   0.0275  &                                                                     \\
  24  &  3s3p($^3$P)3d      &  $^4$D$^o$$_{1/2}$    & 0.0233  &     0.0235     &    0.0237   &   0.0237  &                                                                     \\
  25  &  3s3p($^3$P)3d      &  $^4$P$^o$$_{1/2}$    & 0.0330  &     0.0331     &    0.0354   &   0.0342  &                                                                     \\
  26  &  3s3p($^3$P)3d      &  $^4$P$^o$$_{3/2}$    & 0.0284  &     0.0285     &    0.0316   &   0.0296  &                                                                     \\
  27  &  3s3p($^3$P)3d      &  $^4$D$^o$$_{7/2}$    & 0.0232  &     0.0234     &    0.0244   &   0.0238  &                                                                     \\
  28  &  3s3p($^3$P)3d      &  $^4$D$^o$$_{5/2}$    & 0.0250  &     0.0252     &    0.0261   &   0.0258  &                                                                     \\
  29  &  3s3p($^3$P)3d      &  $^2$D$^o$$_{3/2}$    & 0.0263  &     0.0262     &    0.0274   &   0.0274  &                                                                     \\
  30  &  3s3p($^3$P)3d      &  $^2$D$^o$$_{5/2}$    & 0.0270  &     0.0269     &    0.0282   &   0.0283  &                                                                     \\
  31  &  3s3p($^3$P)3d      &  $^2$F$^o$$_{5/2}$    & 0.0552  &     0.0549     &    0.0575   &   0.0617  &                                                                     \\
  32  &  3s3p($^3$P)3d      &  $^2$F$^o$$_{7/2}$    & 0.0518  &     0.0516     &    0.0544   &   0.0585  &                                                                     \\
  33  &  3s3p($^3$P)3d      &  $^2$P$^o$$_{3/2}$    & 0.0182  &     0.0181     &    0.0185   &   0.0191  &                                                                     \\
  34  &  3s3p($^3$P)3d      &  $^2$P$^o$$_{1/2}$    & 0.0202  &     0.0201     &    0.0200   &   0.0211  &                                                                     \\
  35  &  3s3p($^1$P)3d      &  $^2$F$^o$$_{7/2}$    & 0.0182  &     0.0181     &    0.0190   &   0.0329  &                                                                     \\
  36  &  3s3p($^1$P)3d      &  $^2$F$^o$$_{5/2}$    & 0.0177  &     0.0176     &    0.0185   &   0.0323  &                                                                     \\
  37  &  3s3p($^1$P)3d      &  $^2$P$^o$$_{1/2}$    & 0.0142  &     0.0144     &    0.0152   &   0.0273  &                                                                     \\
  38  &  3s3p($^1$P)3d      &  $^2$D$^o$$_{3/2}$    & 0.0131  &     0.0134     &    0.0154   &   .....   &                                                                     \\
  39  &  3s3p($^1$P)3d      &  $^2$P$^o$$_{3/2}$    & 0.0134  &     0.0133     &    0.0127   &   0.0223  &                                                                     \\
  40  &  3s3p($^1$P)3d      &  $^2$D$^o$$_{5/2}$    & 0.0118  &     0.0119     &    0.0122   &   0.0165  &                                                                     \\
& & & & & & \\ \hline                                                                   
                                            
\end{tabular}                                                                      
}                                                                                  
\begin{flushleft}
{\small
GRASP1: present results with 136 levels from the {\sc grasp} code \\
GRASP2: present results with 332 levels from the {\sc grasp} code \\
CIV3: \cite{gm2} from the {\sc civ3} code  \\ 
MCHF: \cite{cff1} \\ 
a: \cite{mc} \\ 
b: \cite{btp} \\ 
c: \cite{jpl} \\ 
d: \cite{bre} \\ 
e: \cite{et1} \\ 
f: \cite{ehp}, the first entry is for Free M-E Fit, the second for Constrained M-E Fit, 
the 3rd for VNET, and the 4th for ANDC \\  
g: \cite{et2} \\ 
}
\end{flushleft}
\end{minipage}
\end{table*}

In Table 3 we compare our A- values  for some of the E1 transitions, mainly among the lowest 15 levels,  from GRASP1 with those of \cite{sst} and \cite{gyl} from the MCHF and AS codes, respectively. Also included in the table are the f- values from our GRASP1 calculations because they are indicative of the strength of a transition.  As already stated in section 1, Tayal has not reported A- values for all transitions, but for the  ones in common there are no discrepancies with the corresponding results from our calculations. Similarly, our GRASP1 A- values agree within $\sim$ 10\% with those of Liang et al. for almost all transitions,  the only exceptions being a few  weak transitions, such as 8--15 (f $\sim$ 10$^{-5}$) and 12--15 (f $\sim$ 10$^{-6}$). This is because weak(er) transitions are more sensitive to mixing coefficients, and hence differing amount of CI (and methods) often produce different f- values, as discussed in detail by \cite{ah3}. However, in general,  the f-values for weak(er) transitions are less important in comparison to stronger ones with f $\ge$ 0.01, although their inclusion may still be  required in modelling applications.

As in the case of the energy levels, the effect of additional CI on the f- (or A-) values is insignificant  for a majority of the transitions. This is further confirmed by the A- values from our larger GRASP2 calculations with 332 levels, also listed in Table 3. Therefore, the inclusion of CI is important, but only up to a certain extent, mainly when the levels closely interact among themselves, as also discussed earlier by \cite{fe15} for three Mg-like ions. Another useful criterion to assess the accuracy of radiative rates is  the ratio (R) of the velocity and length forms of the S- values, included in Table 3. However, such comparisons are only desirable, and are not a fully sufficient test to assess accuracy, as different calculations (or combinations of configurations) may give comparable f- values in the two forms, but entirely different results in magnitude. Generally, the two forms agree satisfactorily for {\em strong} transitions, but differences  can sometimes be substantial even for some very strong transitions, as demonstrated through various examples by \cite{fe15}. Nevertheless, R is within 0.2 of unity for most (comparatively  strong) transitions, and significant departures are  for only the weaker ones, such as the last 6 listed in Table 3.  Considering all 2733 E1 transitions, R is  within 0.2 of unity for most transitions with f $\ge$ 0.01, but for 70 ($\sim$2.5\%) R is up to a factor of 2. For 38 ($\sim$ 1\%) weaker transitions, R is over 1000 and examples include 7--92 (f $\sim$ 10$^{-10}$) and 77--111  (f $\sim$ 10$^{-11}$). In conclusion, we may confidently state that for a majority of the strong E1 transitions, our radiative rates are accurate to better than 20\%. However, for the weaker transitions this assessment of accuracy does not apply.

\section{Lifetimes}

The lifetime $\tau$ for a level $j$ is defined as follows:

\begin{equation}  {\tau}_j = \frac{1}{{\sum_{i}^{}} A_{ji}}.
\end{equation}
Since this is a measurable parameter, it provides a check on the accuracy of the calculations. Therefore, in Table 1 we have listed our calculated lifetimes, which {\em include}
the contributions from four types of transitions, i.e. E1, E2, M1 and M2. 

In Table 4 we compare our lifetimes with the theoretical results of \cite{gm2} from the  CIV3 code \citep{civ3} and \cite{cff1} from the MCHF code, plus the available measurements of \cite{et1,et2},  \cite{ehp}, \cite{mc},  \cite*{btp}, \cite*{jpl} and \cite{bre}. In general, there are no discrepancies between the two calculations from GRASP and CIV3, or between theory and measurement. However, the differences for the 3s3p($^3$P)3d $^4$F$^o_{3/2}$ level (16) are particularly large. Our lifetime (2.90 ns) and that  (2.11 ns)  calculated by \cite{et2}  are consistent, and comparable to the measured value (1.5 ns). However, the calculated lifetimes by  \cite{gm2} and \cite{cff1}, although agreeing with each other, are lower by up to  a factor of 6. In fact, the  {\em ab initio} calculation of \cite{gm1} gives the lifetime of this level as 1.94 ns, but the use of their adjusted energies and subsequently the A- values lowers this value drastically.

  \section{Collision strengths}

The $R$-matrix radius adopted for  Fe XIV is 4.44 au, and 30  continuum orbitals have been included for each channel angular momentum in the expansion of the wave function. This large expansion is computationally more demanding as the corresponding size of the Hamiltonian matrix is 21,970. However, it  permits us (without any loss of accuracy) to compute $\Omega$ up to an energy of  260 Ryd,  i.e. about 250 Ryd {\em above} the highest threshold of Fe XIV considered in the work. This large range of energy allows us to calculate values of  effective collision strength $\Upsilon$ (see section 6)  up to T$_e$ = 1.2 $\times$10$^{7}$ K, well above  the temperature of maximum abundance in ionisation equilibrium, i.e.  2.0 $\times$10$^{6}$ K  \citep*{pb}.  The maximum number of channels for a partial wave is 728 and  all partial waves with angular momentum $J \le$ 40 are included.  This  large $J$ range  ensures convergence of $\Omega$ for all forbidden and inter-combination transitions, and at all energies. However, for some allowed transitions, particularly at higher energies,   $\Omega$ are not fully converged within this range. We have therefore included a top-up, based on the Coulomb-Bethe  approximation of   \cite{ab}. Furthermore, we have  included such contributions for  forbidden transitions,  based on  geometric series, but these are  small. 

\cite{ps2} have reported values of $\Omega$ for some transitions from the ground state (3s$^2$3p $^2$P$^o_{1/2,3/2}$) to higher excited levels, at an energy of 30 Ryd, and in Table 5 we compare our results with theirs. For the transitions listed in this table, most agree within 10\%, except 2--3/4/5 (i.e. 3s$^2$3p $^2$P$^o_{3/2}$ -- 3s3p$^2$ $^4$P$_{1/2,3/2,5/2}$) where there are differences of up to a factor of 2. These are inter-combination transitions and are forbidden in the $LS$ coupling in which the calculations were (primarily) performed by Storey et al. Since no other results for $\Omega$ are available for comparison purposes,  in Table 6 we list our  $\Omega$ for all resonance transitions and at 6 energies above thresholds. These results will facilitate future comparisons and hence  assessments of  the accuracy of our data. However,  based on comparisons of our calculations for other ions, such as Al X \citep{alx} and Si II \citep{si2}, we assess the accuracy of our listed $\Omega$ to be better than 20\% for most  transitions of Fe XIV.

\begin{table}
 \centering
\begin{minipage}{80mm}
\caption{Comparision of collision strengths ($\Omega$) at an energy of 30 Ryd for some transitions of Fe XIV. } 
\begin{tabular}{rrll}                                                                                                                                                                                                                                                                                           
\hline                                                                                                                                                          
$i$ & $j$ & Present Results & \cite{ps2} \\ 
\hline                                                                                                                                                              
1  &   2  & 0.176  &  ......   \\
1  &   3  & 0.0171 &  0.0165   \\
1  &   4  & 0.0064 &  0.0062   \\
1  &   5  & 0.0050 &  0.0047   \\
1  &   6  & 0.888  &  0.931    \\
1  &   7  & 0.0142 &  0.0141   \\
1  &   8  & 1.675  &  1.701    \\
1  &   9  & 1.078  &  1.166    \\
1  &  10  & 1.106  &  1.162    \\
1  &  11  & 2.767  &  2.918    \\
1  &  12  & 0.0281 &  0.0293   \\
2  &   3  & 0.0050 &  0.0080   \\
2  &   4  & 0.0056 &  0.0141   \\
2  &   5  & 0.0299 &  0.0512   \\
2  &   6  & 0.0556 &  0.0617   \\
2  &   7  & 1.327  &  1.395    \\
2  &   8  & 0.151  &  0.170    \\
2  &   9  & 1.926  &  1.954    \\
2  &  10  & 5.656  &  5.835    \\
2  &  11  & 0.742  &  0.764    \\
2  &  12  & 5.249  &  5.449    \\
\hline                                                                                  
                                                                           
\end{tabular}                                                                           
\end{minipage}                                                                           
\end{table}

\begin{table*}      
\caption{Collision strengths for the resonance transitions of  Fe XIV. ($a{\pm}b \equiv$ $a\times$10$^{{\pm}b}$).  (For complete table see Supporting Information.)}          
\begin{tabular}{rrlllllll}                                                                                    
\hline                                                                                                        
\multicolumn{2}{c}{Transition} & \multicolumn{6}{c}{Energy (Ryd)}\\                                           
\hline                                                                                                        
  $i$ & $j$ &    20  & 50 & 100 & 150 & 200 & 250 \\                                                          
\hline     
  1 &  2 &  1.800$-$01 &  1.733$-$01 &  1.717$-$01 &  1.713$-$01 &  1.708$-$01 &  1.702$-$01 \\
  1 &  3 &  1.783$-$02 &  1.665$-$02 &  1.658$-$02 &  1.614$-$02 &  1.560$-$02 &  1.497$-$02 \\
  1 &  4 &  8.892$-$03 &  3.907$-$03 &  1.870$-$03 &  1.256$-$03 &  9.793$-$04 &  8.327$-$04 \\
  1 &  5 &  7.032$-$03 &  2.862$-$03 &  1.156$-$03 &  6.600$-$04 &  4.501$-$04 &  3.420$-$04 \\
  1 &  6 &  8.135$-$01 &  9.974$-$01 &  1.138$-$00 &  1.168$-$00 &  1.147$-$00 &  1.137$-$00 \\
  1 &  7 &  1.782$-$02 &  1.069$-$02 &  8.329$-$03 &  7.915$-$03 &  7.865$-$03 &  7.903$-$03 \\
  1 &  8 &  1.523$-$00 &  1.894$-$00 &  2.249$-$00 &  2.379$-$00 &  2.383$-$00 &  2.345$-$00 \\
  1 &  9 &  9.804$-$01 &  1.220$-$00 &  1.455$-$00 &  1.555$-$00 &  1.567$-$00 &  1.544$-$00 \\
  1 & 10 &  1.005$-$00 &  1.252$-$00 &  1.484$-$00 &  1.581$-$00 &  1.586$-$00 &  1.594$-$00 \\
   .  &    .    &  . & . & . & . & . &  . &  \\ 
    .  &    .    &  . & . & . & . & . &  . &  \\     
        .  &    .    &  . & . & . & . & . &  . &  \\ 
 \hline                                                                                                        
\end{tabular}                                                                                                 
\end{table*}                                                                                                  

\begin{figure*}
\includegraphics[angle=90,width=0.9\textwidth]{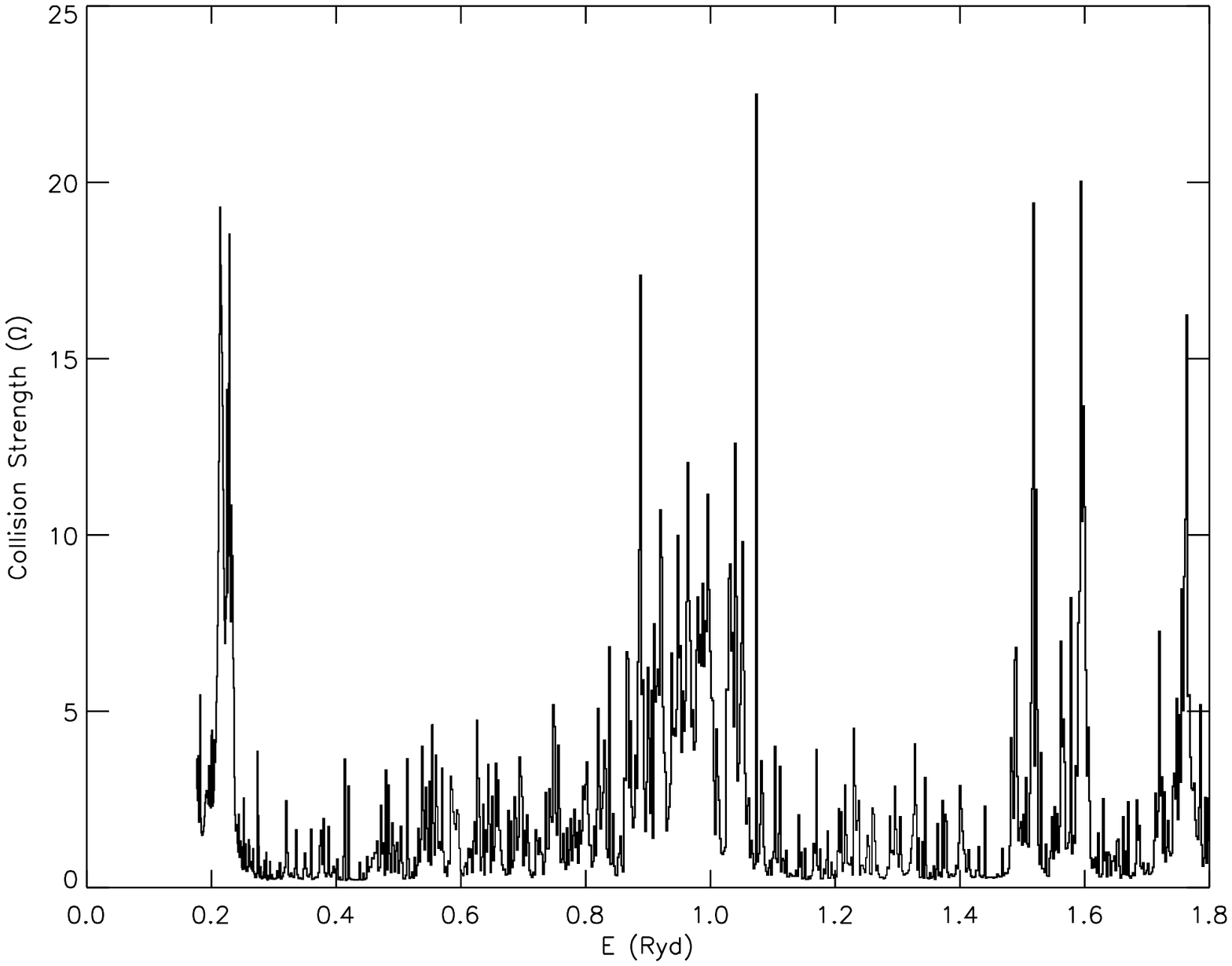}
 \caption{Collision strengths for the 3s$^2$3p $^2$P$^o_{1/2}$ -- 3s$^2$3p $^2$P$^o_{3/2}$ (1--2) transition of Fe XIV.}
\end{figure*}

\newpage
\clearpage
\begin{figure*}
 \includegraphics[angle=90,width=0.9\textwidth]{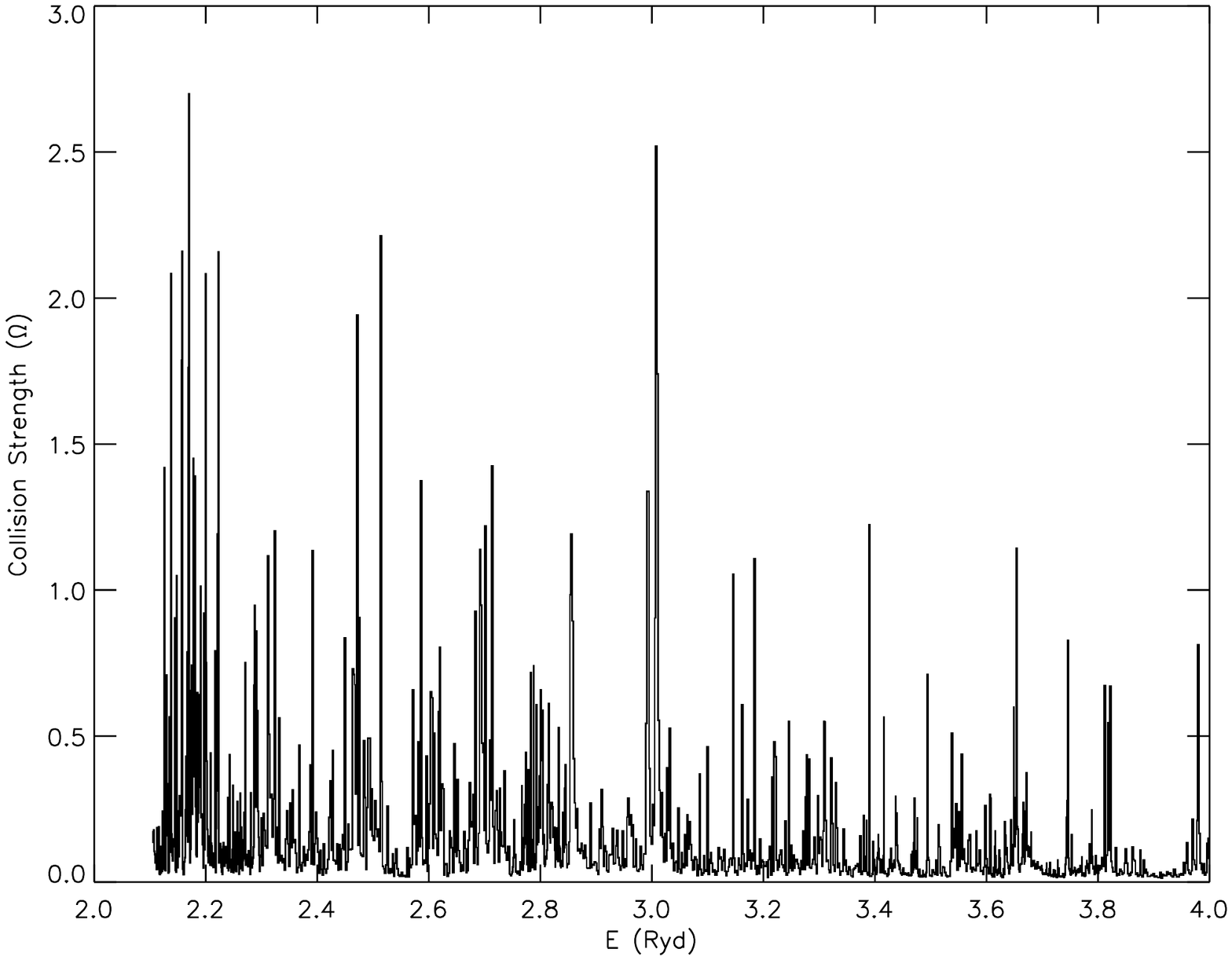}
 \caption{Collision strengths for the 3s$^2$3p $^2$P$^o_{1/2}$ -- 3s$^2$3p $^2$P$^o_{1/2}$ -- 3s3p$^2$ $^4$P$_{3/2}$ (1--4) transition of Fe XIV.}
\end{figure*}

\begin{figure*}
 \includegraphics[angle=90,width=0.9\textwidth]{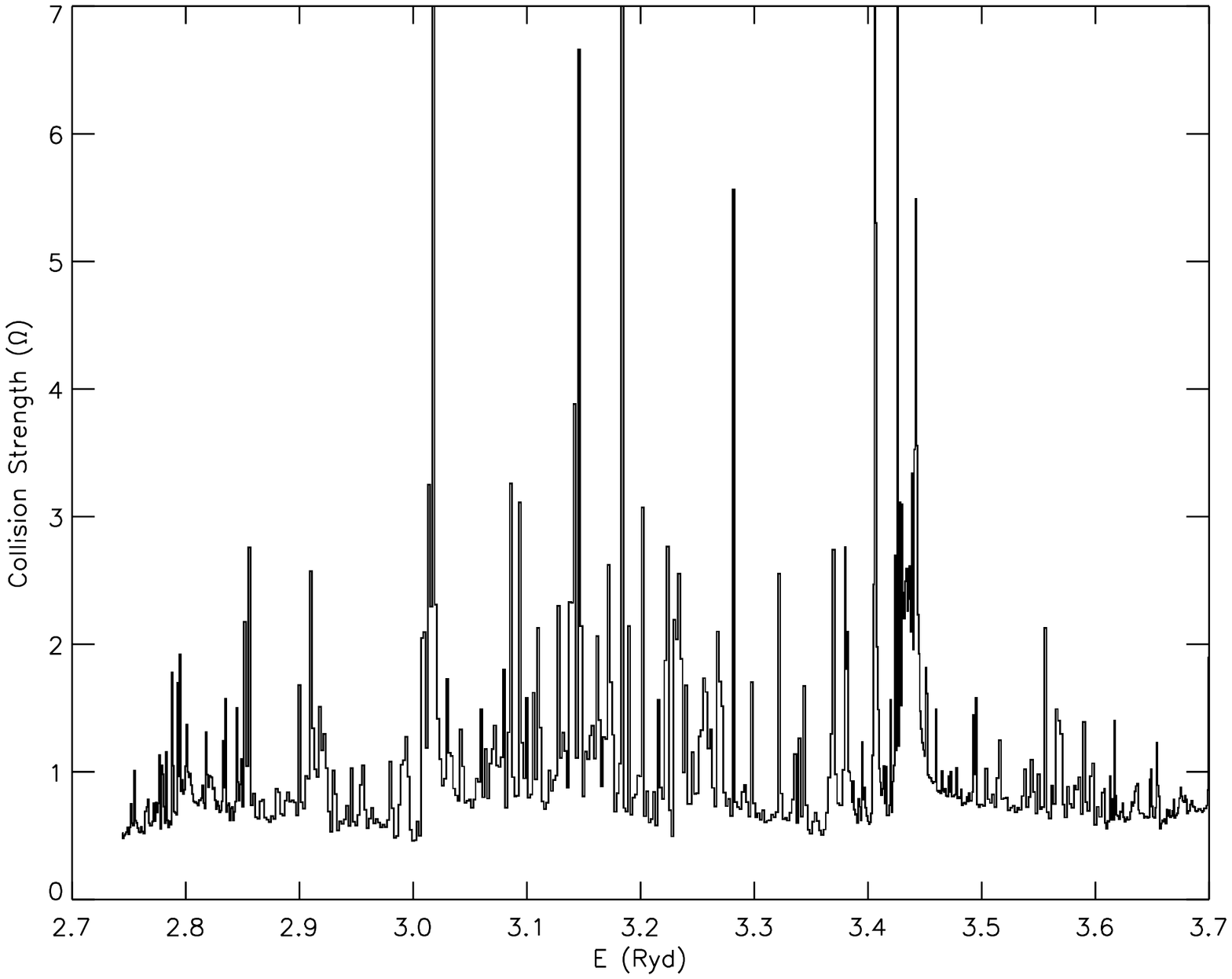}
 \caption{Collision strengths for the 3s$^2$3p $^2$P$^o_{1/2}$ -- 3s$^2$3p $^2$P$^o_{1/2}$ -- 3s3p$^2$ $^2$D$_{3/2}$ (1--6) transition of Fe XIV.}
\end{figure*}

\begin{figure*}
\includegraphics[angle=-90,width=0.9\textwidth]{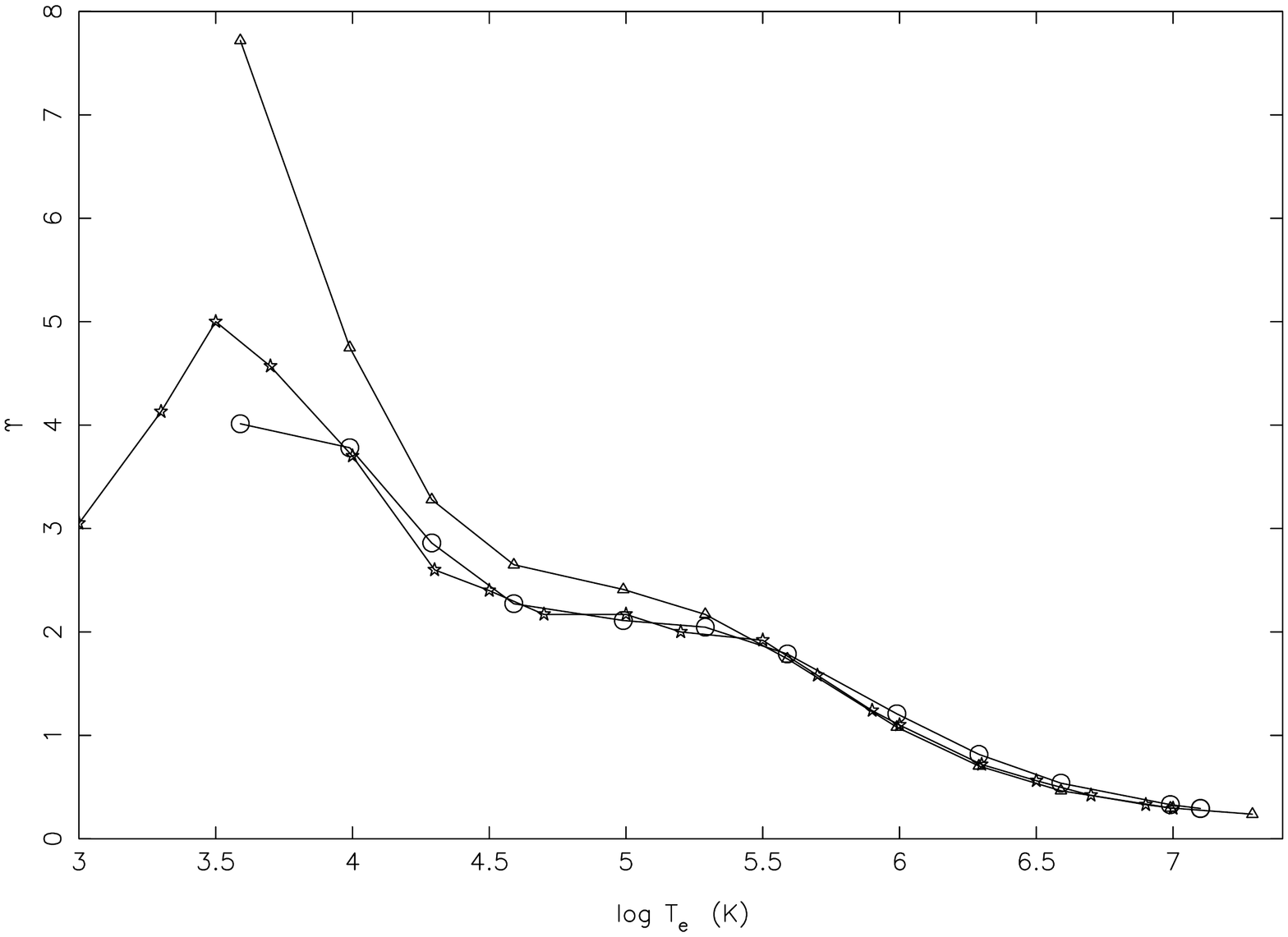}
 \caption{Comparison of effective collision strengths for the 3s$^2$3p $^2$P$^o_{1/2}$ -- 3s$^2$3p $^2$P$^o_{3/2}$ (1-- 2) transition of Fe XIV. Circles: present results, 
 stars:  Tayal (2008) and triangles:  Liang et al. (2010). }
\end{figure*}

\begin{figure*}
\includegraphics[angle=90,width=0.9\textwidth]{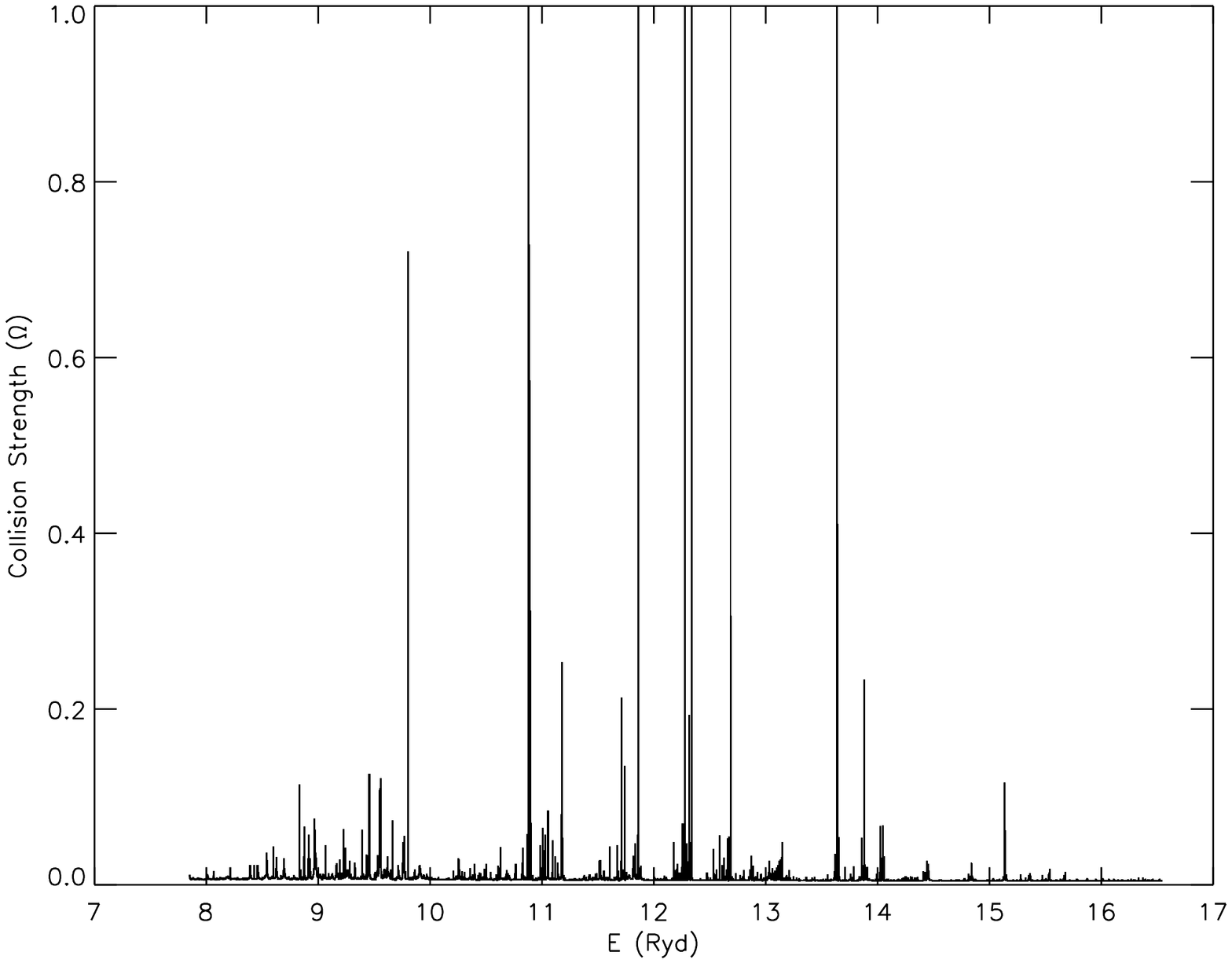}
 \caption{Collision strengths for the 3s$^2$3p $^2$P$^o_{1/2}$ --  3s3p($^1$P)3d $^2$P$^o_{3/2}$ (1-- 39) transition of Fe XIV.}
\end{figure*}

\section{Effective collision strengths}

As already shown by \cite{sst} and \cite{gyl}, $\Omega$ does not vary smoothly within  the thresholds region,  especially for (semi) forbidden transitions. Such  closed-channel (Feshbach) resonances need to be resolved  in a fine energy mesh  to accurately account for their contribution. In most astrophysical  plasmas, electrons have a velocity  distribution and therefore an averaged value, known as {\em effective} collision strength ($\Upsilon$) is required. The most commonly used velocity distribution is {\em Maxwellian}, i.e. 

\begin{equation}
\Upsilon(T_e) = \int_{0}^{\infty} {\Omega}(E) \, {\rm exp}(-E_j/kT_e) \,d(E_j/{kT_e}),
\end{equation}
where $k$ is  Boltzmann's constant, T$_e$  electron temperature in K, and E$_j$  the electron energy with respect to the final (excited) state. Once the value of $\Upsilon$ is
known the corresponding results for the excitation q(i,j) and de-excitation q(j,i) rates can be easily obtained from the following equations:

\begin{equation}
q(i,j) = \frac{8.63 \times 10^{-6}}{{\omega_i}{T_e^{1/2}}} \Upsilon \, {\rm exp}(-E_{ij}/{kT_e}) \hspace*{1.0 cm}{\rm cm^3s^{-1}}
\end{equation}
and
\begin{equation}
q(j,i) = \frac{8.63 \times 10^{-6}}{{\omega_j}{T_e^{1/2}}} \Upsilon \hspace*{1.0 cm}{\rm cm^3 s^{-1}},
\end{equation}
where $\omega_i$ and $\omega_j$ are the statistical weights of the initial ($i$) and final ($j$) states, respectively, and E$_{ij}$ is the transition energy. The contribution of resonances may greatly enhance the values of $\Upsilon$ over those of the background  collision strengths ($\Omega_B$), but this depends strongly on the type of transition,   and particularly the temperature.  In general, the lower the T$_e$, the greater is  the contribution of resonances. However,  values of $\Omega$ need to  be calculated over a wide energy range (above thresholds)  to obtain convergence of the integral in Eq. (7), as demonstrated in fig. 7 of  \cite{ni11}. For this reason we have calculated values of $\Omega$ up to an energy of 260 Ryd, as discussed in section 5 and hence in our work there is no need to invoke the high energy formulations of \cite{bt92},  as undertaken by \cite{gyl}.

To resolve resonances, we have  calculated $\Omega$ in (most of) the thresholds region in a narrow energy mesh  of 0.001 Ryd, although a mesh of 0.002 Ryd has also been used whenever the energy difference between any two thresholds is wider, such as levels 2 and 3. In total $\Omega$ have been calculated  at over  10,500 energies, and in Figs. 1 -- 3 we show  resonance structure for three transitions, namely   1 -- 2 (3s$^2$3p $^2$P$^o_{1/2}$ -- 3s$^2$3p $^2$P$^o_{3/2}$), 1 -- 4 (3s$^2$3p $^2$P$^o_{1/2}$ -- 3s3p$^2$ $^4$P$_{3/2}$) and 1 -- 6 (3s$^2$3p $^2$P$^o_{1/2}$ -- 3s3p$^2$ $^2$D$_{3/2}$), which are forbidden, inter-combination and allowed, respectively. Similar resonances in  these transitions  are also apparent from the data of  \cite{ps1} and  \cite{sst} -- see figs. 1--3 of the latter.  For  the important 1--2 transition, the magnitude of the resonances shown by \cite{ps1} are higher by almost a factor of 2, and this has also been noted by \cite{gyl} -- see their fig. 3. As a consequence of this and some other factors, the $\Upsilon$ calculated by \cite{ps1} are significantly larger (by over a factor of 5), particularly towards the lower end of the temperature range, as demonstrated in fig. 4 of \cite{sst}.  Resonances in our calculations for this transition are closer in magnitude to those of \cite{sst}, but are comparatively greater in number. Similarly for the 1 -- 4 transition, the resonance structure in our calculations is denser but their magnitude is more comparable to that of \cite{sst} than of \cite{ps1}.  The density of resonances for the 1 -- 6 transition is even more apparent in our work than that of \cite{sst}, although the peaks are narrow in width and hence may not make a significant contribution to  $\Upsilon$.

\begin{table*}                                                                                                                              
\caption{Effective collision strengths for transitions in  Fe XIV. ($a{\pm}b \equiv a{\times}10^{{\pm}b}$). (For complete table see Supporting Information.)}                                
\begin{tabular}{rrlllllllll}                                                                                                                
\hline                                                                                                                                      
\multicolumn {2}{c}{Transition} & \multicolumn{8}{c}{Temperature (log T$_e$, K)}\\                                                          
\hline                                                                                                                                      
$i$ & $j$ &    5.70 &  5.90  &    6.10  &    6.30   &   6.50  &    6.70  &    6.90  &   7.10    \\                                         
\hline                                                                                                                                                 
  1 &  2 &  1.643$-$00 &  1.344$-$00 &  1.055$-$00 &  8.065$-$01 &  6.115$-$01 &  4.677$-$01 &  3.660$-$01 &  2.925$-$01 \\
  1 &  3 &  6.773$-$02 &  5.311$-$02 &  4.181$-$02 &  3.358$-$02 &  2.783$-$02 &  2.392$-$02 &  2.118$-$02 &  1.886$-$02 \\
  1 &  4 &  9.967$-$02 &  7.390$-$02 &  5.375$-$02 &  3.868$-$02 &  2.764$-$02 &  1.964$-$02 &  1.387$-$02 &  9.712$-$03 \\
  1 &  5 &  1.059$-$01 &  7.747$-$02 &  5.553$-$02 &  3.929$-$02 &  2.755$-$02 &  1.916$-$02 &  1.322$-$02 &  9.024$-$03 \\
  1 &  6 &  1.487$-$00 &  1.323$-$00 &  1.181$-$00 &  1.080$-$00 &  1.025$-$00 &  1.011$-$00 &  1.019$-$00 &  1.010$-$00 \\
  1 &  7 &  2.148$-$01 &  1.633$-$01 &  1.221$-$01 &  8.988$-$02 &  6.542$-$02 &  4.739$-$02 &  3.446$-$02 &  2.524$-$02 \\
  1 &  8 &  1.287$-$00 &  1.307$-$00 &  1.350$-$00 &  1.420$-$00 &  1.519$-$00 &  1.643$-$00 &  1.770$-$00 &  1.834$-$00 \\
  1 &  9 &  8.417$-$01 &  8.575$-$01 &  8.903$-$01 &  9.398$-$01 &  1.003$-$00 &  1.080$-$00 &  1.158$-$00 &  1.197$-$00 \\
  1 & 10 &  8.847$-$01 &  8.952$-$01 &  9.201$-$01 &  9.621$-$01 &  1.023$-$00 &  1.102$-$00 &  1.186$-$00 &  1.229$-$00 \\
    .  &    .    &  . & . & . & . & . & . & .  &  .  \\ 
    .  &    .    &  . & . & . & . & . & . & .  &  .  \\     
        .  &    .    &  . & . & . & . & . & . & .  &  .  \\ 

\hline                                                                                                                                      
\end{tabular}                                                                                                                               
\end{table*}             

Our calculated values of $\Upsilon$ are listed in Table 7 over a wide temperature range  up to log T$_e$ =  7.1 K, suitable for applications to a wide range  of  astrophysical plasmas. The inaccuracy of the $\Upsilon$ of \cite{ps2} has already been discussed and demonstrated by both \cite{sst} and \cite{gyl}, and therefore we make no comparison with their results.  However, before we discuss any comparisons with the calculations of Tayal, we note that this is not straightforward,  because his energy levels, A- values and  $\Upsilon$ in tables 1, 3/4 and 5, respectively, have different orderings. The problem has partly arisen because his energy levels are neither in the  NIST ordering nor in increasing energy, and mainly because he used different sets of wavefuntions for atomic structure (i.e. to determine energy levels and A- values) and the scattering process, i.e. to calculate $\Omega$ and $\Upsilon$. Therefore, for undertaking a comparison with the Tayal data  a careful (re)ordering is required for  all his 135 levels, although his results for $\Upsilon$ are only provided up to level 59. 

 For the most important transition of Fe XIV, i.e.  (3s$^2$3p) $^2$P$^o_{1/2 }$ -- $^2$P$^o_{3/2}$ (1 -- 2), our  results are shown in Fig. 4 along with those of \cite{sst} and \cite{gyl}. The $\Upsilon$ of Tayal are comparable to our calculations (within $\sim$20\%), but those of Liang et al. are significantly higher (by up to a factor of two), particularly towards the lower end of the temperature range. In fact, at the lowest common temperature of 10$^{3.593}$ K, the $\Upsilon$ of Liang et al. differ with our calculations by over 20\% for $\sim$70\% of the transitions, and in a  majority of cases their results are higher. For some transitions, such as 1 -- 62/66/70/82/88/91/92/93, their data are greater by over an order of magnitude. Some differences between  two calculations are understandable at low temperatures, because of the position of resonances, but not such large discrepancies as noted here. Similarly, at the highest common temperature of 10$^{6.991}$ K, the discrepancies between our results and those of Liang et al. are over 20\% for $\sim$46\% of the transitions, and as at lower temperatures their $\Upsilon$ values  are higher for almost all transitions. Discrepancies of up a factor of 2 are common for many transitions, but are  over an order of magnitude for some, such as 1 -- 94/112/131, 2 -- 94/132 and 3 -- 83/106/122. The most likely reasons for such large discrepancies are those stated in section 1, but an error in the version of the code adopted by \cite{gyl}  cannot be ruled out. Some versions of the $R$- matrix code are known to contain errors, particularly related to the Breit-Pauli part which accounts for the relativistic effects -- see, for example, \cite{fexv} who found large  differences in $R$- matrix calculations for transitions in Fe XV, as did  \cite{li1,li2}  more recently for Li-like ions. Another (strong) possibility is the presence of pseudo resonances, because in the generation of wave functions   \cite{gyl} have included levels of additional 77 configurations which are not spectroscopic -- see section 2. If the pseudo resonances are not properly smoothed over then the calculated values of $\Upsilon$ may be abnormally high.  The ICFT results calculated by \cite{gyl} can sometimes  significantly overestimate the values of $\Upsilon$, in comparison to the  corresponding Breit-Pauli work, at least for some of the transitions, as recently demonstrated and discussed by \cite*{ps3}, see in particular their fig. 3. Finally, the Liang et al. data contain massive resonances for the 1 -- 39 (3s$^2$3p $^2$P$^o_{1/2 }$ -- 3s3p($^1$P)3d $^2$P$^o_{3/2}$) transition of Fe XIV (see their fig. 5) at energies below $\sim$10 Ryd. However, our calculations do not show this, as may be seen from  Fig. 5. As a result,   the $\Upsilon$ of Liang et al. for this transition are larger than the present calculations and those of  Tayal by a factor  of 2.

\begin{table*}                                                                                                                                                  
\caption{Comparision of effective collision strengths ($\Upsilon$) at three temperatures   for  transitions among the lowest 10 levels of Fe XIV.  $a{\pm}b \equiv a{\times}$10$^{{\pm}b}$.}                           
\begin{tabular}{rrlllllllll}                                                                                                                                   
\hline                                                                                                                                                                                                                                                                                                                 
 \multicolumn{2}{c}{Transition}  & \multicolumn{3}{c}{Present Results} & \multicolumn{3}{c}{\cite{gyl}} & \multicolumn{3}{c}{\cite{sst}}  \\     
 \hline                                                                                                              
$i$ & $j$  & 1.0$\times$10$^6$ K & 2.0$\times$10$^6$ K & 1.0$\times$10$^7$ K & 1.0$\times$10$^6$ K & 2.0$\times$10$^6$ K & 1.0$\times$10$^7$ K & 1.0$\times$10$^6$ K & 2.0$\times$10$^6$ K & 1.0$\times$10$^7$ K \\
\hline                                                                                                                                                                                                                            
    1  &  2 & 1.209$-$0  &  8.065$-$1 & 3.299$-$1  &    1.08$-$0 & 7.04$-$1 & 2.98$-$1  &  1.10$+$0 & 7.18$-$1 &  2.94$-$1 \\
    1  &  3 & 4.755$-$2  &  3.358$-$2 & 2.012$-$2  &    4.87$-$2 & 3.47$-$2 & 2.06$-$2  &  5.50$-$2 & 3.70$-$2 &  2.00$-$2 \\
    1  &  4 & 6.403$-$2  &  3.868$-$2 & 1.180$-$2  &    6.29$-$2 & 3.99$-$2 & 1.25$-$2  &  6.70$-$2 & 4.10$-$2 &  1.20$-$2 \\
    1  &  5 & 6.669$-$2  &  3.929$-$2 & 1.112$-$2  &    7.18$-$2 & 4.40$-$2 & 1.25$-$2  &  6.90$-$2 & 4.10$-$2 &  1.10$-$2 \\
    1  &  6 & 1.254$-$0  &  1.080$-$0 & 1.020$-$0  &    8.37$-$1 & 8.41$-$1 & 1.02$-$0  &  7.91$-$1 & 7.90$-$1 &  9.65$-$1 \\
    1  &  7 & 1.433$-$1  &  8.988$-$2 & 2.988$-$2  &    1.34$-$1 & 8.43$-$2 & 2.82$-$2  &  1.31$-$1 & 8.10$-$2 &  2.60$-$2 \\
    1  &  8 & 1.323$-$0  &  1.420$-$0 & 1.813$-$0  &    1.30$-$0 & 1.40$-$0 & 1.84$-$0  &  1.25$+$0 & 1.33$+$0 &  1.77$+$0 \\
    1  &  9 & 8.702$-$1  &  9.398$-$1 & 1.184$-$0  &    8.87$-$1 & 9.52$-$1 & 1.25$-$0  &  1.00$+$0 & 1.06$+$0 &  1.35$+$0 \\
    1  & 10 & 9.046$-$1  &  9.621$-$1 & 1.215$-$0  &    9.15$-$1 & 9.68$-$1 & 1.25$-$0  &  9.44$-$1 & 9.82$-$1 &  1.23$+$0 \\
    2  &  3 & 5.133$-$2  &  3.062$-$2 & 1.149$-$2  &    5.17$-$2 & 3.29$-$2 & 1.39$-$2  &  5.70$-$2 & 3.50$-$2 &  1.40$-$2 \\
    2  &  4 & 1.115$-$1  &  6.569$-$2 & 2.049$-$2  &    1.10$-$1 & 6.99$-$2 & 2.58$-$2  &  1.22$-$1 & 7.40$-$2 &  2.50$-$2 \\
    2  &  5 & 1.990$-$1  &  1.273$-$1 & 5.698$-$2  &    1.97$-$1 & 1.36$-$1 & 7.13$-$2  &  2.03$-$1 & 1.33$-$1 &  6.50$-$2 \\
    2  &  6 & 3.140$-$1  &  2.081$-$1 & 9.169$-$2  &    2.33$-$1 & 1.61$-$1 & 8.77$-$2  &  2.22$-$1 & 1.50$-$1 &  8.30$-$2 \\
    2  &  7 & 1.441$-$0  &  1.363$-$0 & 1.448$-$0  &    1.42$-$0 & 1.37$-$0 & 1.57$-$0  &  1.31$+$0 & 1.26$+$0 &  1.51$+$0 \\
    2  &  8 & 1.717$-$1  &  1.582$-$1 & 1.679$-$1  &    1.98$-$1 & 1.85$-$1 & 2.04$-$1  &  2.31$-$1 & 2.17$-$1 &  2.54$-$1 \\
    2  &  9 & 1.536$-$0  &  1.644$-$0 & 2.078$-$0  &    1.55$-$0 & 1.65$-$0 & 2.16$-$0  &  1.54$+$0 & 1.62$+$0 &  2.11$+$0 \\
    2  & 10 & 4.496$-$0  &  4.828$-$0 & 6.144$-$0  &    4.52$-$0 & 4.85$-$0 & 6.36$-$0  &  4.63$+$0 & 4.90$+$0 &  6.29$+$0 \\
    3  &  4 & 4.055$-$1  &  2.514$-$1 & 8.283$-$2  &    4.59$-$1 & 2.92$-$1 & 9.37$-$2  &  4.21$-$1 & 2.64$-$1 &  8.50$-$2 \\
    3  &  5 & 3.796$-$1  &  2.959$-$1 & 2.105$-$1  &    4.68$-$1 & 3.55$-$1 & 2.28$-$1  &  3.96$-$1 & 3.10$-$1 &  2.12$-$1 \\
    3  &  6 & 9.266$-$2  &  5.558$-$2 & 1.675$-$2  &    8.80$-$2 & 5.62$-$2 & 1.71$-$2  &  8.30$-$2 & 5.20$-$2 &  1.60$-$2 \\
    3  &  7 & 6.618$-$2  &  3.818$-$2 & 9.992$-$3  &    6.94$-$2 & 4.27$-$2 & 1.13$-$2  &  6.60$-$2 & 4.00$-$2 &  1.10$-$2 \\
    3  &  8 & 3.067$-$2  &  1.747$-$2 & 4.315$-$3  &    3.94$-$2 & 2.36$-$2 & 5.89$-$3  &  3.40$-$2 & 2.10$-$2 &  5.00$-$3 \\
    3  &  9 & 2.157$-$2  &  1.177$-$2 & 2.668$-$3  &    3.01$-$2 & 1.77$-$2 & 4.35$-$3  &  5.00$-$2 & 2.70$-$2 &  6.00$-$3 \\
    3  & 10 & 3.244$-$2  &  1.814$-$2 & 4.598$-$3  &    4.37$-$2 & 2.55$-$2 & 6.29$-$3  &  3.70$-$2 & 2.20$-$2 &  5.00$-$3 \\
    4  &  5 & 1.604$-$0  &  1.067$-$0 & 4.548$-$1  &    1.35$-$0 & 9.15$-$1 & 4.19$-$1  &  1.26$+$0 & 8.59$-$1 &  4.02$-$1 \\
    4  &  6 & 1.588$-$1  &  9.461$-$2 & 2.572$-$2  &    1.36$-$1 & 8.63$-$2 & 2.46$-$2  &  1.32$-$1 & 8.20$-$2 &  2.50$-$2 \\
    4  &  7 & 2.740$-$1  &  1.614$-$1 & 4.662$-$2  &    2.13$-$1 & 1.34$-$1 & 4.12$-$2  &  2.10$-$1 & 1.31$-$1 &  4.00$-$2 \\
    4  &  8 & 5.016$-$2  &  2.854$-$2 & 7.051$-$3  &    5.16$-$2 & 3.14$-$2 & 8.06$-$3  &  5.20$-$2 & 3.20$-$2 &  8.00$-$3 \\
    4  &  9 & 4.891$-$2  &  2.707$-$2 & 6.688$-$3  &    4.70$-$2 & 2.81$-$2 & 7.31$-$3  &  4.80$-$2 & 2.80$-$2 &  7.00$-$3 \\
    4  & 10 & 1.001$-$1  &  5.503$-$2 & 1.312$-$2  &    9.85$-$2 & 5.76$-$2 & 1.44$-$2  &  9.90$-$2 & 5.80$-$2 &  1.50$-$2 \\
    5  &  6 & 1.524$-$1  &  9.259$-$2 & 2.915$-$2  &    1.52$-$1 & 9.72$-$2 & 3.07$-$2  &  1.39$-$1 & 8.80$-$2 &  2.70$-$2 \\
    5  &  7 & 3.491$-$1  &  2.211$-$1 & 8.114$-$2  &    3.80$-$1 & 2.48$-$1 & 8.76$-$2  &  3.19$-$1 & 2.11$-$1 &  7.30$-$2 \\
    5  &  8 & 5.292$-$2  &  3.169$-$2 & 9.999$-$3  &    6.16$-$2 & 3.82$-$2 & 1.09$-$2  &  5.50$-$2 & 3.40$-$2 &  1.00$-$2 \\
    5  &  9 & 5.833$-$2  &  3.418$-$2 & 1.012$-$2  &    5.28$-$2 & 3.25$-$2 & 9.08$-$3  &  5.20$-$2 & 3.20$-$2 &  9.00$-$3 \\
    5  & 10 & 1.307$-$1  &  7.161$-$2 & 1.683$-$2  &    1.40$-$1 & 8.31$-$2 & 2.10$-$2  &  1.33$-$1 & 7.90$-$2 &  2.00$-$2 \\
    6  &  7 & 9.828$-$1  &  6.233$-$1 & 2.586$-$1  &    9.64$-$1 & 6.20$-$1 & 2.60$-$1  &  9.00$-$1 & 5.82$-$1 &  2.43$-$1 \\
    6  &  8 & 2.869$-$1  &  2.353$-$1 & 1.885$-$1  &    2.90$-$1 & 2.42$-$1 & 1.96$-$1  &  2.93$-$1 & 2.50$-$1 &  2.05$-$1 \\
    6  &  9 & 2.639$-$1  &  1.705$-$1 & 8.318$-$2  &    2.51$-$1 & 1.66$-$1 & 8.03$-$2  &  2.50$-$1 & 1.64$-$1 &  7.59$-$2 \\
    6  & 10 & 2.587$-$1  &  1.539$-$1 & 5.095$-$2  &    2.65$-$1 & 1.62$-$1 & 5.24$-$2  &  2.61$-$1 & 1.59$-$1 &  5.27$-$2 \\
    7  &  8 & 4.262$-$1  &  3.665$-$1 & 3.128$-$1  &    4.31$-$1 & 3.74$-$1 & 3.21$-$1  &  4.41$-$1 & 3.88$-$1 &  3.44$-$1 \\
    7  &  9 & 1.592$-$1  &  1.138$-$1 & 6.808$-$2  &    1.62$-$1 & 1.18$-$1 & 6.89$-$2  &  1.51$-$1 & 1.09$-$1 &  5.66$-$2 \\
    7  & 10 & 4.992$-$1  &  2.802$-$1 & 7.187$-$2  &    4.74$-$1 & 2.76$-$1 & 7.14$-$2  &  4.36$-$1 & 2.54$-$1 &  6.79$-$2 \\
    8  &  9 & 2.385$-$1  &  1.377$-$1 & 3.462$-$2  &    2.82$-$1 & 1.66$-$1 & 4.17$-$2  &  2.75$-$1 & 1.64$-$1 &  4.11$-$2 \\
    8  & 10 & 3.933$-$1  &  2.325$-$1 & 7.232$-$2  &    4.48$-$1 & 2.65$-$1 & 7.67$-$2  &  4.12$-$1 & 2.44$-$1 &  7.00$-$2 \\
    9  & 10 & 3.550$-$1  &  2.563$-$1 & 1.541$-$1  &    5.02$-$1 & 3.48$-$1 & 1.81$-$1  &  3.70$-$1 & 2.71$-$1 &  1.66$-$1 \\
\hline                                                                                
\end{tabular}                                                 
\end{table*}

In Table 8, we compare our results of $\Upsilon$ with those of  \cite{gyl} and  \cite{sst}  for transitions among the lowest 10 levels of Fe XIV,  at three  common temperatures of 1.0$\times$10$^6$, 2.0$\times$10$^6$ and 1.0$\times$10$^7$ K. For these transitions at these temperatures, all sets of $\Upsilon$ (generally) agree within $\sim$20\%, although the differences are slightly larger for a few, such as 2 -- 6, 3 -- 9/10 and 9  --10. Such  good agreement among three independent calculations is highly satisfactory and encouraging, but is somewhat contrary to the discussion above. Furthermore, the most relevant temperature for Fe XIV is 2.0$\times$10$^6$ K, that of maximum abundance in ionisation equilibrium \citep{pb}.  At this T$_e$ the discrepancies are less than 20\% between our calculated $\Upsilon$ and those of  \cite{sst} for all transitions. However, this comparison is possible for only 630 transitions ( (i.e. $\sim$7\% of the total), because of the limited data reported by Tayal. Unfortunately, a similar comparison made at this T$_e$ with the $\Upsilon$ of  \cite{gyl} reveals that for about half the transitions there are  differences of  over 20\%, and their results are mostly higher. Therefore, based on the comparisons shown in Table 8 and Fig. 4 and discussed above, we believe that the values of  $\Upsilon$ calculated by Liang et al. are overestimated for almost all transitions and at all temperatures. Hence, a re-examination of their results is  desirable.

\section{Conclusions}

In this paper we have presented results for energy levels and  radiative rates for four types of transitions (E1, E2, M1 and M2) among the lowest 136 levels of Fe XIV.
Additionally, lifetimes of all the levels have been reported, although  measurements are available for only a few, for which there is good ageeement between theory and
experiment.  Based on a variety of comparisons, our energy levels are assessed to be accurate to better than 1\%, and the results for radiative rates, oscillator strengths, line
strengths and lifetimes are assessed to be accurate to better than 20\% for a majority of the strong transitions (levels). Similarly, the accuracy of our data for collision
strengths and effective collision strengths is estimated to be better than 20\% for a majority of  transitions. The earlier calculations of \cite{gyl} appear to have overestimated the values of $\Upsilon$ for almost all transitions and over an entire  range of temperature.  We believe the present set of  results for radiative and excitation rates  are  the most complete and reliable  to date, and hence should be useful for the modelling of a variety of plasmas. Finally,  as $\Upsilon$ is a slowly varying function of T$_e$,  corresponding data at any other temperature within the reported range can be easily interpolated,  or may be requested from  the first author. 

\section*{Acknowledgments}

KMA is thankful to  the Atomic Weapons Establishment, Aldermaston for    financial support.

\begin{flushleft} 
{\bf SUPPORTING INFORMATION}\\
Additional Supporting Information may be found in the online version of this article:\\
{\bf Table 2.} Transition wavelengths ($\lambda_{ij}$ in $\rm \AA$), radiative rates (A$_{ji}$ in s$^{-1}$), oscillator strengths (f$_{ij}$, dimensionless), and line     
strengths (S, in atomic units) for electric dipole (E1), and A$_{ji}$ for E2, M1 and M2 transitions in Fe XIV. ($a{\pm}b \equiv a{\times}$10$^{{\pm}b}$). \\
{\bf Table 6.} Collision strengths for the resonance transitions of    Fe XIV. ($a{\pm}b \equiv$ $a\times$10$^{{\pm}b}$). \\
{\bf Table 7.} Effective collision strengths for transitions in  Fe XIV. ($a{\pm}b \equiv a{\times}10^{{\pm}b}$).
\end{flushleft}


\begin{thebibliography}{99}
\bibitem[\protect\citeauthoryear{Aggarwal \& Keenan}{2012}]{li1}
Aggarwal K. M.,  Keenan F. P., 2012, At. Data Nucl. Data Tables, 98, 1003
\bibitem[\protect\citeauthoryear{Aggarwal \& Keenan}{2013}]{li2}
Aggarwal K. M.,  Keenan F. P., 2013, At. Data Nucl. Data Tables, 99, 156
\bibitem[\protect\citeauthoryear{Aggarwal \& Keenan}{2014a}]{alx}
Aggarwal K. M.,  Keenan F. P., 2014a, MNRAS, 438, 1223
\bibitem[\protect\citeauthoryear{Aggarwal \& Keenan}{2014b}]{si2}
Aggarwal K. M.,  Keenan F. P., 2014b, MNRAS, 442, 388
\bibitem[\protect\citeauthoryear{Aggarwal et al.}{2000}]{fexv}
Aggarwal K. M., Deb N. C., Keenan F. P., Msezane A. Z., 2000, J. Phys. B, 33, L391
\bibitem[\protect\citeauthoryear{Aggarwal \& Keenan}{2008}]{ni11}
Aggarwal K. M.,  Keenan F. P., 2008,  Eur. Phys. J.,  D 46,  205 
 \bibitem[\protect\citeauthoryear{Aggarwal et al.}{2007}]{fe15}
Aggarwal  K. M., Tayal V., Gupta G. P.,   Keenan F. P., 2007,  At. Data Nucl. Data Tables, 93, 615  
\bibitem[\protect\citeauthoryear{Badnell}{1997}]{as}
Badnell N. R.,  1997, J. Phys. B, 30, 1
\bibitem[\protect\citeauthoryear{Beiersdorfer, Tr{\" a}bert  \& Pinnington}{Beiersdorfer et al.}{2003}]{btp}
Beiersdorfer P., Tr{\" a}bert E.,   Pinnington E. H.,  2003, ApJ, 587, 836
\bibitem[\protect\citeauthoryear{Berrington et al.}{1978}]{rm1}
Berrington K. A., Burke P. G., LeDourneuf M., Robb W. D., Taylor K. T.,  Vo Ky Lan,  1978, Comput. Phys. Commun.,  14, 367
\bibitem[\protect\citeauthoryear{Berrington, Eissner \& Norrington}{Berrington et al.}{1995}]{rm2}
Berrington K. A., Eissner W. B., Norrington P. H., 1995,	Comput. Phys. Commun.,  92,  290  
\bibitem[\protect\citeauthoryear{Bhatia et al.}{1994}]{akb}
Bhatia A. K., Kastner S. O., Keenan F. P., Conlon E. S., Widing K. G., 1994, ApJ, 427, 497
\bibitem[\protect\citeauthoryear{Brenner et al.}{2007}]{bre}
Brenner G., Crespo L{\' o}pez-Urrutia J. R., Harman Z., Mokler P. H.,  Ullrich J., 2007, Phys. Rev., A 75, 032504
\bibitem[\protect\citeauthoryear{Brickhouse, Raymond \& Smith}{Brickhouse et al.}{1995}]{nsb}
Brickhouse N. S., Raymond J. C.,  Smith B. W. 1995, ApJS, 97, 551
\bibitem[\protect\citeauthoryear{Brosius, Davila \& Thomas}{Brosius et al.}{1998}]{jwb}
Brosius J. W., Davila J. M.,  Thomas R. J., 1998, ApJ, 497, L113
\bibitem[\protect\citeauthoryear{Brown et al.}{2008}]{cmb}
Brown C. M., Feldman U., Seely J. F., Korendyke C. M.,  Hara H.,  2008,  ApJS, 176, 511
\bibitem[\protect\citeauthoryear{Bryans, Landi \& Savin}{Bryans et al.}{2009}]{pb}
Bryans P., Landi E.,    Savin D. W., 2009,  ApJ,  691, 1540 
\bibitem[\protect\citeauthoryear{Burgess \& Sheorey}{1974}]{ab}
Burgess A.,   Sheorey V. B.,  1974, J. Phys.,  B7,  2403 
\bibitem[\protect\citeauthoryear{Burgess \& Tully}{1992}]{bt92}
Burgess A.,    Tully J. A., 1992, A\&A,   254, 436 
\bibitem[\protect\citeauthoryear{Del Zanna, Berrington \& Mason}{Del Zanna et al.}{2004}]{del04}
Del Zanna G., Berrington K. A.,  Mason H.E.,  2004, Astron. Astrophys. 422, 731
\bibitem[\protect\citeauthoryear{Dermendjiev, Kolarov, \& Mitsev}{Dermendjiev et al.}{1992}]{spl}
Dermendjiev V. N., Kolarov G. V.,  Mitsev Ts. A.,  1992, Sol. Phys., 137, 199
\bibitem[\protect\citeauthoryear{Dong et al.}{2006}]{czd}
Dong C. Z., Kato T., Fritsche S.,  Koike F.,  2006 MNRAS, 369, 1735
\bibitem[\protect\citeauthoryear{Dufton \& Kingston}{1991}]{dk}
Dufton P. L., Kingston A. E., 1991, Phys. Scr., 43, 386
\bibitem[\protect\citeauthoryear{Eissner, Jones \& Nussbaumer}{Eissner et al.}{1974}]{ss}
Eissner W., Jones M.,  Nussbaumer H.,  1974, Comput. Phys.  Commun., 8, 270
\bibitem[\protect\citeauthoryear{Ferguson, Korista \& Ferland}{Ferguson et al.}{1997}]{jwf}
Ferguson J. W., Korista K. T.,  Ferland G. J.,  1997, ApJS, 110, 287
\bibitem[\protect\citeauthoryear{Fisher}{1978}]{rrf}
Fisher R. R.,  1978, Sol. Phys., 57, 119
\bibitem[\protect\citeauthoryear{Froese-Fischer}{1991}]{mchf}
Froese-Fischer C.,  1991, Comput. Phys. Commun.,  64, 369
\bibitem[\protect\citeauthoryear{Froese-Fischer, Tachiev \& Irimia}{Froese-Fischer et al.}{2006}]{cff1}
Froese-Fischer C.,  Tachiev G.,   Irimia A.,  2006, At. Data Nucl. Data Tables,  92, 607
\bibitem[\protect\citeauthoryear{Grant et al.}{1980}]{grasp0}
Grant I. P., McKenzie B. J.,  Norrington P. H.,  Mayers D. F.,   Pyper N. C., 1980,	Comput. Phys. Commun.,  21,  207  
\bibitem[\protect\citeauthoryear{Gupta \& Msezane}{2001}]{gm1}
Gupta G. P.,  Msezane A. Z.,  2001, J. Phys. B 34, 4217
\bibitem[\protect\citeauthoryear{Gupta \& Msezane}{2005}]{gm2}
Gupta G. P.,  Msezane A. Z.,  2005, At. Data Nucl. Data Tables,  89, 1 
\bibitem[\protect\citeauthoryear{Hibbert}{1975}]{civ3}
Hibbert A., 1975, Comput. Phys. Commun.,  9,  141  
\bibitem[\protect\citeauthoryear{Hibbert}{2000}]{ah3}
Hibbert A.,  2000, AIP Conf. Proc., 543, 242
\bibitem[\protect\citeauthoryear{Huang}{1989}]{knh}
Huang K. N., 1989, At. Data Nucl. Data Tables, 34, 1
\bibitem[\protect\citeauthoryear{Jup{\' e}n,  Isler  \& Tr{\" a}bert}{Jup{\' e}n et al.}{1993}]{jup}
Jup{\' e}n C., Isler R. C.,  Tr{\" a}bert E., 1993, MNRAS, 264, 627
\bibitem[\protect\citeauthoryear{Kramida et al.}{2013}]{nist}
Kramida A., Ralchenko Y., Reader J. and NIST ASD Team,  2013, NIST Atomic Spectra Database (version 5.1) Online at: http://physics.nist.gov/asd
\bibitem[\protect\citeauthoryear{Lamzin, Stempels \& Piskunov}{Lamzin et al.}{2001}]{lam}
Lamzin S. A., Stempels H. C.,  Piskunov N. E., 2001, A\&A, 369, 965
\bibitem[\protect\citeauthoryear{Liang et al.}{2010}]{gyl}
Liang G. Y., Badnell N. R., Crespo L{\' o}pez-Urrutia J. R., Baumann T. M., Del Zanna G., Storey P. J., Tawara H., Ullrich J., 2010, ApJS, 190, 322
\bibitem[\protect\citeauthoryear{Moehs \& Church}{1999}]{mc}
Moehs D. P.,  Church D. A., 1999, ApJ, 516, L111
\bibitem[\protect\citeauthoryear{Pinnington et al.}{1990}]{ehp}
 Pinnington E. H., Ansbacher W., Tauheed T., Tr{\" a}bert E.,  Heckmann P. H., M{\" o}ller G.,  Blanke J. H., 1990, Z. Phys. D 17, 5
\bibitem[\protect\citeauthoryear{Safronova et al.}{2002}]{uis}
Safronova U. I., Sataka M., Albritton J. R., Johnson W. R.,  Safronova M. S., 2002, Phys. Rev. A, 65, 022507
\bibitem[\protect\citeauthoryear{Smith, Chutjian \& Lozano}{Smith et al.}{2005}]{jpl}
Smith S. J., Chutjian A., Lozano J. A., 2005, Phys. Rev., A 72, 062504
\bibitem[\protect\citeauthoryear{Storey, Mason \& Saraph}{Storey et al.}{1996}]{ps1}
Storey P. J.,  Mason H. E.,  Saraph H. E., 1996, A\&A, 309, 677
\bibitem[\protect\citeauthoryear{Storey, Mason \& Young}{Storey et al.}{2000}]{ps2}
Storey P. J.,  Mason H. E.,  Young P. R., 2000, A\&AS, 141, 285
\bibitem[\protect\citeauthoryear{Storey, Sochi \& Badnell}{Storey et al.}{2014}]{ps3}
Storey P. J.,  Sochi T.,  Badnell N. R., 2014, MNRAS, 441, 3028
\bibitem[\protect\citeauthoryear{Tayal}{2008}]{sst}
Tayal S. S., 2008, ApJS, 178, 359
\bibitem[\protect\citeauthoryear{Thomas \& Neupert}{1994}]{tn}
Thomas R. J.,  Neupert W. M., 1994, ApJS, 91, 461
\bibitem[\protect\citeauthoryear{Tr{\" a}bert et al.}{1988}]{et1}
Tr{\" a}bert E.,  Heckmann P. H., Hutton R.,  Martinson I., 1988, J. Opt. Soc. Am. B 5, 2173
\bibitem[\protect\citeauthoryear{Tr{\" a}bert et al.}{1993}]{et2}
Tr{\" a}bert E., Wagner C., Heckmann P. H.,  M{\" o}ller G.,  Brage T., 1993, Phys. Scr., 48, 593
\bibitem[\protect\citeauthoryear{Wei et al.}{2008}]{hlw}
Wei H. L., Zhang H., Ma C. W., Zhang J. Y.,  Cheng X. L., 2008, Phys. Scr., 77, 035301
\bibitem[\protect\citeauthoryear{Young, Landi \& Thomas}{Young et al.}{1998}]{pry}
Young P. R., Landi E.,  Thomas R. J., 1998, A\&A, 329, 291
\bibitem[\protect\citeauthoryear{Zatsarinny \& Froese-Fischer}{1999}]{zff}
Zatsarinny O., Froese-Fischer C., 1999, Comput. Phys. Commun.,  124,  247  


\end{thebibliography}
\end{document}